\documentclass[aps,pra,twocolumn,showpacs,superscriptaddress,longbibliography]{revtex4-2}

\usepackage[T1]{fontenc}
\usepackage{graphicx}
\usepackage{dcolumn}
\usepackage{mathtools} 
\usepackage{bm,amssymb}
\usepackage{braket}
\usepackage[normalem]{ulem}
\usepackage{xcolor} 
\usepackage[most]{tcolorbox}
\usepackage{booktabs}
\usepackage{diagbox}
\usepackage{soul}

\usepackage[hidelinks]{hyperref}
\usepackage{orcidlink}

\graphicspath{{figures/}}

\makeatletter
\renewcommand\thesection{\arabic{section}}
\renewcommand\section{\@startsection
  {section}{1}{\z@}%
  {-2.0ex \@plus -0.8ex \@minus -0.2ex}%
  {0.9ex \@plus 0.2ex}%
  {\normalfont\bfseries\raggedright}}
 \renewcommand\subsection{\@startsection
  {subsection}{2}{\z@}%
  {-1.6ex \@plus -0.6ex \@minus -0.2ex}%
  {0.6ex \@plus 0.2ex}%
  {\normalfont\bfseries\raggedright}}
\makeatother

\begin{document}

\title{Self-calibrated multiparameter measurement of three-dimensional microwave fields}

\author{Yupeng Wang\orcidlink{0009-0005-4409-4402}}%
\thanks{These authors contributed equally to this work.}
\affiliation{ 
Department of Physics and Astronomy, Purdue University, West Lafayette, IN 47907, USA
}%

\author{Xinghan Wang\orcidlink{0009-0006-0313-7952}}%
\thanks{These authors contributed equally to this work.}
\affiliation{ 
Department of Physics and Astronomy, Purdue University, West Lafayette, IN 47907, USA
}%

   \author{Aishik Panja\orcidlink{0009-0003-4794-7736}}
\thanks{These authors contributed equally to this work.}
\affiliation{ 
Department of Physics and Astronomy, Purdue University, West Lafayette, IN 47907, USA
}

\author{Md. Ehsanuzzaman}%
\affiliation{ 
Department of Physics and Astronomy, Purdue University, West Lafayette, IN 47907, USA
}%

\author{Chuan-Hsun Li}%
\affiliation{ 
Department of Physics and Astronomy, Purdue University, West Lafayette, IN 47907, USA
}%

\author{Qi-Yu Liang\orcidlink{0000-0002-2430-7248}}
\affiliation{ 
Department of Physics and Astronomy, Purdue University, West Lafayette, IN 47907, USA
}%
\affiliation{Purdue Quantum Science and Engineering Institute, Purdue University, West Lafayette, IN 47907, USA}

\date{\today}

\begin{abstract}
Rydberg atoms are promising for microwave (MW) sensing and control, but full local MW characterization remains difficult. Existing methods generally do not provide self-calibrated reconstruction of the three-dimensional vector field, which is valuable for both atom-based sensing and in-situ field characterization in complex electromagnetic environments. We propose and implement multi-level, Zeeman-resolved Rydberg electromagnetically induced transparency (EIT) spectroscopy in a laser-cooled atomic ensemble. We extract the three polarization amplitudes from a single spectrum and show that the MW polarization components give rise to closed interferometric loops within the atoms' internal Hilbert space, enabling extraction of their relative phases.
Moreover, it is self-calibrated and requires no external reference MW fields, with MW parameters largely separable from one another and from other experimental parameters. These features make it broadly applicable to dedicated sensing platforms as well as quantum optics and quantum information experiments.
\end{abstract}

\maketitle
\section{Introduction}
Rydberg atomic sensors provide sensitive and versatile means for detecting electromagnetic fields over a wide frequency range, from near-DC to terahertz regimes~\cite{adams2019rydberg,kitching2025atom,chandra2026electrometry,wang2026sensing,Hammerland2026,glick2026low}. In most implementations, however, the measurement is limited to the field amplitude or its projection along a single axis~\cite{yuan2023quantum,liu2023electric}. In many applications, one needs the full local microwave (MW) field, including its polarization and even the relative phase between different polarization components. This need is relevant, for example, in atom-based MW sensing and in the use of MW fields for control in quantum optics and quantum information experiments.

In the sensing context, three-dimensional (3D) vector polarimetry offers richer information, including the ability to determine the wave vector (k-vector or angle-of-arrival)~\cite{talashila2025determining,elgee2025electrically}. Recent works have made significant progress, but typically assume the MW or radio-frequency field is linearly polarized and retrieve only the orientation angle relative to a reference axis, such as the laser polarization in all-optical schemes~\cite{sedlacek2013atom,jiao2017atom,cloutman2025quantum} or the local oscillator (LO) in heterodyne configurations~\cite{wang2023precise}. A very recent all-optical, self-calibrated approach~\cite{chilcott2026quantum} extended Rydberg-atom polarimetry to elliptically polarized MW fields and demonstrated extraction of a phase characterizing the ellipticity. A complete 3D characterization of the field, without assuming a known angle-of-arrival, requires three amplitudes and two relative phases. Its full reconstruction has been demonstrated using heterodyne detection with three mutually orthogonal LOs~\cite{elgee2024complete}, but this approach requires detailed calibration of the LO amplitudes and their relative phases.

In cold-atom quantum optics and quantum information experiments, MW dressing can significantly reduce sensitivity of Rydberg atoms to DC stray electric fields~\cite{bohorquez2023reducing} and tune atom-atom interactions~\cite{palm2026enhanced,kurdak2025enhancement,young2021asymmetric,sevinccli2014microwave,shi2017annulled}, enabling tunable interaction strength and sign, as well as engineered three-body terms. Such interaction engineering requires precise control over polarization. However, in typical cold-atom experiments not designed specifically for MW sensing, the in-vacuo components, the vacuum system, along with components surrounding it create a complex electromagnetic environment. As a result, modeling and achieving pure and controllable polarization is challenging, making in-situ Rydberg-atom electrometry an appealing solution. Recent work~\cite{kurdak2025high} separated Zeeman levels with large magnetic fields to measure each polarization component at a different frequency and developed a procedure to control the polarization.

Here, we propose and experimentally demonstrate a self-calibrated method for 3D MW field measurement via Zeeman-sublevel spectroscopy. We implement this scheme in a laser-cooled atomic platform, which is relevant both to MW-enabled control in quantum optics experiments and to atom-based sensing. Although the above-mentioned sensing works are all conducted with vapor cells, cold-atom platforms~\cite{TuSQLelectrometer,duverger2024metrology,jamieson2025continuous,liao2020microwave,zhou2023improving,zhang2025microwave,wang2026quantum} are of growing interest for atom-based sensing because they suppress Doppler effects that limits vapor-cell sensitivity and also offer the potential to harness additional quantum effects~\cite{nill2024avalanche,wang2026robust,chen2024quantum,shu2024eliminating,chen2026background} for enhanced sensing capabilities.

Our approach extracts all polarization components at a single MW frequency. This feature is especially advantageous in situations where frequency scans are impractical or where the MW polarization varies over tens of megahertz. The scheme is phase sensitive, enabling extraction of one or both relative phases between the $\sigma^+$, $\sigma^-$ and $\pi$ components, depending on the number of available interferometric loops. This method complements existing microwave electrometry techniques that rely on calibration of external reference MW fields. Because our apparatus is designed for quantum optics rather than sensing, the present demonstration emphasizes the broad applicability of the method across experimental platforms while making clear that the measurement capability demonstrated here does not reflect the full potential of the approach in a sensing-optimized apparatus.

\section{Results}
\subsection{Physical model}

\begin{figure*}[ht]
\centering
\includegraphics[width=0.85\textwidth]{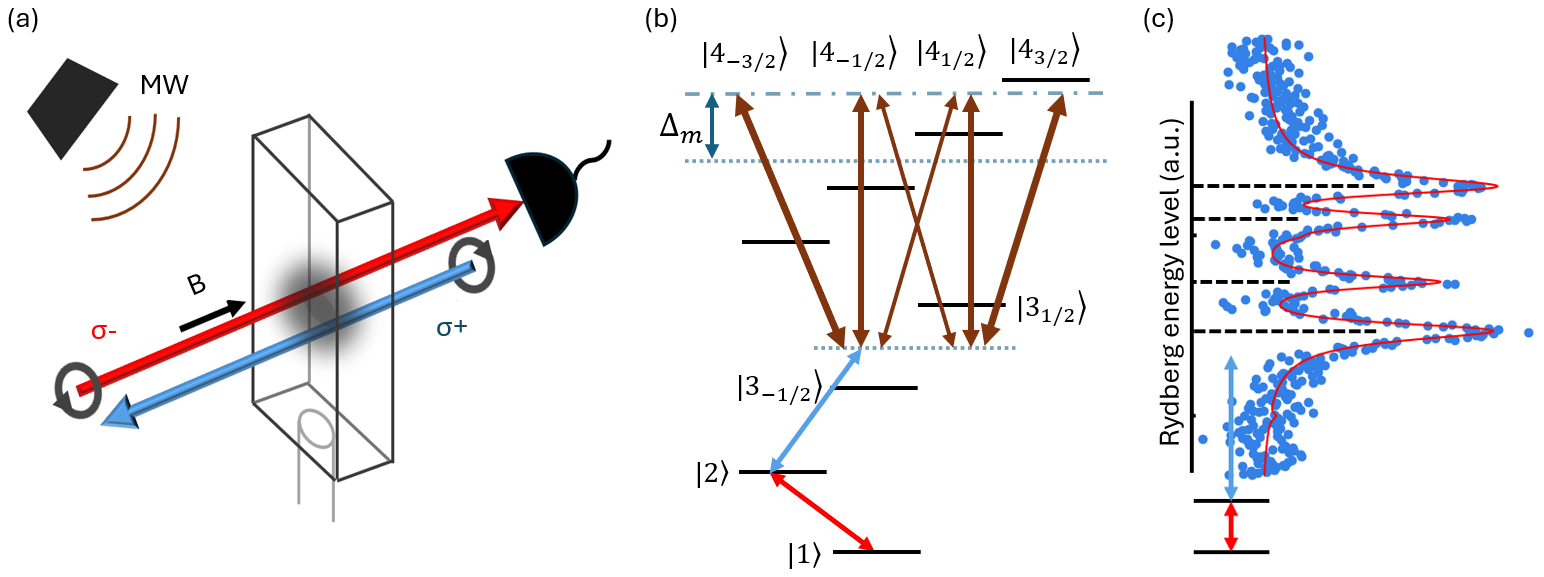}
\caption{Scheme for 3D microwave field measurement. (a) Experimental setup showing counter-propagating probe ($\sigma^-$) and control ($\sigma^+$) beams passing through a cold atomic cloud in a vacuum chamber, with a bias magnetic field B. (b) Energy level diagram. Red, blue, and brown arrows denote the probe, control, and microwave fields, respectively. The lower and upper dotted lines indicate the zero-field degenerate energy levels of $\ket{3_{m_j}}$ and $\ket{4_{m_j}}$, respectively. $\Delta_m$ denotes the MW detuning from the $\ket{3_{m_j}} \leftrightarrow \ket{4_{m_j'}}$ transition. (c) Representative probe-transmission spectrum (blue dots: data; red line: fit) together with the corresponding energy levels. The ground and intermediate states are shown as solid horizontal lines, while dashed lines indicate the calculated eigenenergies of the MW-dressed Rydberg states. The length of each dashed line is proportional to the projection (magnitude squared) of the corresponding eigenstate onto $\ket{3_{-1/2}}$.}
\label{fig:setup & levels & spectrum}
\end{figure*}

We consider Rydberg electromagnetically induced transparency (EIT) in a spin-polarized cloud of $^{87}$Rb atoms. The atoms are initially prepared in the ground state $\ket{1} = \ket{5S_{1/2}, F=2, m_F=-2}$, where $F$ and $m_F$ denote the hyperfine and magnetic sublevels, respectively. As shown in Fig.~\ref{fig:setup & levels & spectrum}(a), a weak $\sigma^-$-polarized probe laser couples the ground state to an intermediate state $\ket{2} = \ket{5P_{3/2}, F=3, m_F=-3}$, which is coupled to a Rydberg state $\ket{3_{-1/2}} =\ket{61S_{1/2},m_j=-1/2}$ via a $\sigma^+$-polarized control laser. Additionally, a MW field addresses the $\ket{3_{-1/2}} \leftrightarrow \ket{4_{m_j}}=\ket{61P_{3/2}, m_j }$ transition (Fig.~\ref{fig:setup & levels & spectrum}(b)). Here, $j$ and $m_j$ denote the total angular momentum quantum number, and its projection along the quantization axis, respectively.

In the absence of external fields, the Hamiltonian of the system is given by $H=H_0+H_l+H_{m}$, where $H_0$ is the atom energy in rotating-frame, $H_l$ represents the probe and control couplings, and $H_m$ is the MW couplings:
\begin{subequations}
    \begin{align}
    H_0 = 
 & - \hbar \bigg[\Delta_p\ket{2}\bra{2} + \sum_{{m_j}=-1/2}^{1/2}(\Delta_p+\Delta_c)\ket{3_{m_j}}\bra{3_{m_j}}\nonumber\\& + \sum_{m_j=-3/2}^{3/2}(\Delta_p+\Delta_c+\Delta_m)\ket{4_{m_j}}\bra{4_{m_j}}\bigg]
 \\
H_l = & \hbar \bigg[\frac{\Omega_p}{2}\ket{1}\bra{2} + \frac{\Omega_c}{2}\ket{2}\bra{3_{-1/2}} + \text{h.c.} \bigg]
\\
    H_m = & \hbar  \sum_{m_j=-1/2}^{1/2}\sum_{q=-1}^{1}\frac{\Omega_m^{m_j,q}}{2}\ket{3_{m_j}}\bra{4_{m_j+q}} + \text{h.c.} 
\end{align}
\label{eq:Hamiltonian}
\end{subequations}
``h.c.''~denotes the Hermitian conjugate.
$\Delta_m$ is the MW detuning from the $\ket{3_{m_j}} \leftrightarrow \ket{4_{m_j'}}$ transition.
$\Delta_p$ and $\Delta_c$ are the probe and control laser detunings from the $\ket{1} \leftrightarrow \ket{2}$ and $\ket{2} \leftrightarrow \ket{3_{-1/2}}$ transitions, respectively, while 
$\Omega_p$ and $\Omega_c$ are their corresponding Rabi frequencies. 
Note that the detunings are defined relative to the zero-field transition frequencies, where the Zeeman sublevels are degenerate in the absence of applied bias fields.

$\Omega_m^{m_j,q}$ is the MW Rabi frequency for polarization index $q$, with $q=+1$ ($\sigma^+$), $q=0$ ($\pi$) and $q=-1$ ($\sigma^-$). $\Omega_m^{-1/2,q}$ and $\Omega_m^{1/2,q}$ correspond to the same MW field and are related by their respective Clebsch-Gordan coefficients. They are defined as:
\begin{align}
    \Omega_m^{m_j,q} =  e \textbf{E}_q \cdot \bra{3_{m_j}} \textbf{r} \ket{4_{m_j+q}}  e^{i\phi_q}
\end{align}
where $e$ is the elementary charge, $\mathbf{E}_q$ is the
q spherical component of the microwave electric field, $-e\langle 3_{m_j}| \mathbf{r} |4_{m_j+q}\rangle$ is the transition dipole moment, and $\phi_q$ is the phase of the $q$ polarization. The probe and control phases can be absorbed into the definitions of the atomic states and therefore do not affect any physical prediction. The phases $\phi_q$ contain one redundant degree of freedom: a common shift of all three $\phi_q$ corresponds only to a phase convention.

In the presence of DC magnetic and electric fields, the total Hamiltonian includes additional terms describing Zeeman splittings of magnetic sublevels and DC Stark shifts of Rydberg states, i.e. $H_{\text{tot}}=H+H_B+H_E$.
\begin{subequations}
    \begin{align}
    H_B &= \mu_B B \bigg(\sum_{i=1,2} g_F^i m_F^i\ket{i}\bra{i} 
    + \sum_{\substack{i=3,4\\ m_j}} g_j^i m_j \ket{i_{m_j}}\bra{i_{m_j}} \bigg)
    \\
    H_E &= -\sum_{\substack{i=3,4\\ m_j}}\frac{1}{2} \alpha_{i,m_j} E_{\text{DC}}^2
\end{align}
\end{subequations}
where $\mu_B$ is the Bohr magneton, $B$ is the magnetic field, $g_F^i$ are the hyperfine Landé g-factors for states $i=1,2$, $g_j^i$ are the fine-structure Landé g-factors for Rydberg states $i=3,4$, $E_{\text{DC}}$ is the DC electric field, and $\alpha_{i,m_j}$ is the scalar polarizability of state $\ket{i_{m_j}}$.

To account for decay and dephasing processes, we describe the system dynamics using the Lindblad master equation $
    \dot{\rho}=\frac{i}{\hbar}[\rho,H_{\text{tot}}]   +\mathcal{L}(\rho) $ for the system's density matrix $\rho$, with the Lindblad superoperator:

\begin{equation}
   \mathcal{L}(\rho)=-\sum_{k}\frac{1}{2}\left(  L_k^\dagger L_k\rho  + \rho L_k^\dagger L_k  -2L_k\rho L_k^\dagger \right)
   \label{eq:Liouvillian operator}
\end{equation}
$L_2=\sqrt{\Gamma_2}\ket{1}\bra{2}$, together with $L_k=\sqrt{\Gamma_k}\ket{k}\bra{k}$, $k\in\{3,4\}$, are the only Lindbladian terms considered. Both the 61S$_{1/2}$ and 61P$_{3/2}$ Rydberg states have lifetime longer than $100~\mu$s at room temperature, resulting in negligible decay rates compared to other decoherence processes. $\Gamma_2/(2\pi)=6.1~$MHz is the spontaneous decay rate of the intermediate state $\ket{2}$.

The steady-state solution of the density matrix is associated with the probe transmission $T$ through:
\begin{equation}
    T=\exp\left[\text{OD Im}(\rho_{21})\frac{\Gamma_2}{\Omega_p}\right]
    \label{eq: transmission}
\end{equation}
where OD denotes the optical depth.

\subsection{Spectra interpretation}
 We consider probe transmission spectra as functions of the control detuning $\Delta_c$, with the probe kept resonant with the Zeeman-shifted $|1\rangle \leftrightarrow |2\rangle$ transition, resulting in a low baseline transmission.
With the MW dressing, the resulting spectrum can be understood as EIT with six possible third states in the ladder, rather than just $\ket{3_{-1/2}}$. As each MW-dressed state is brought into EIT resonance by scanning $\Delta_c$, a peak in probe transmission is expected (Fig.~\ref{fig:setup & levels & spectrum}(c)). However, $\ket{3_{-1/2}}$ is directly MW-coupled to three Zeeman sublevels ($\ket{4_{-3/2}}$, $\ket{4_{-1/2}}$ and $\ket{4_{1/2}}$), forming a strongly mixed four-state subspace. Consequently, four of the six dressed states have a much larger population in $\ket{3_{-1/2}}$, the only Rydberg state addressed by the control laser. In practice, the number and visibility of observed peaks depend on the Zeeman shifts, DC Stark shifts, MW detuning and amplitudes of different polarization components, the probe and control Rabi frequencies, decoherence, and experimental noise.

\subsection{Parameter extraction}
In a multi-level spectrum, many parameters can be extracted from the same fit, but their roles depend on the measurement objective.
Aside from atomic-structure parameters and fundamental constants, the relevant model parameters include the MW polarization amplitudes and phase, $E_{\text{DC}}$, $B$, $\Omega_c$, $\Gamma_{3}$, $\Gamma_{4}$, OD, and the probe and control resonance frequencies.
We refer to the quantities of interest as \textit{sensing} parameters and to the quantities that set the spectroscopic conditions as \textit{control} parameters. One likely objective is to extract MW parameters, although one could instead extract other quantities, such as the DC electric field and magnetic field. 

The parameters exhibit a degree of separability, arising from their predominantly distinct spectroscopic signatures:
OD sets the transmission baseline, while a small residual probe detuning relative to the Zeeman-shifted $|1\rangle \leftrightarrow |2\rangle$ resonance ($\lesssim500~$kHz in our experiments) introduces a baseline asymmetry along the scanned $\Delta_c$ axis. The magnetic field $B$ contributes a distinct Zeeman splitting pattern, and a small offset in the control resonance frequency produces an overall shift of the spectrum along the $\Delta_c$ axis. $\Gamma_{3}$,  $\Gamma_{4}$ and $\Omega_c$ each act through a single global parameter value that shapes all peaks simultaneously, while each MW polarization amplitude and phase predominantly affects a different subset of peaks. These distinct mechanisms largely decouple the sensing parameters from the control parameters, while keeping correlations among the sensing parameters themselves relatively weak. Because of this separability, we can either fix well-calibrated parameters to reduce fitting time or allow most relevant parameters to vary freely in the fit. Calibration is achieved through auxiliary spectroscopic measurements within the same atomic system, so the method remains self-calibrated and does not require external reference MW fields.

 The experimental probe-transmission spectra are fitted using the model described by Eqs.~(\ref{eq:Hamiltonian}-\ref{eq: transmission}), extended to include two additional nearby Rydberg states. The off-resonant couplings become relevant under some experimental conditions. The fits are performed using a nonlinear least-squares routine (\texttt{curve\_fit} in \texttt{scipy.optimize}). Detailed fitting procedures and complete fitted parameter sets for all spectra in this manuscript are provided in Supplement 1. 
 In our analysis, we treat OD, $B$, $E_{\text{DC}}$, $\Gamma_{3}$ and $\Gamma_{4}$ as fit parameters, with all other control parameters independently determined. 
In Fig.~\ref{fig:setup & levels & spectrum}(c), the MW-dressed states are calculated from the parameters determined by the fit result (red line).
The eigenenergies of these dressed states match the observed peak positions in the spectrum while their overlap with $\ket{3_{-1/2}}$ qualitatively accounts for the relative peak heights.

\subsection{Experiment}

We now describe the experimental conditions under which the spectra are obtained. We emphasize that our setup is not specifically designed for MW sensing. Nevertheless, this method is applicable to a wide range of cold-atom platforms and can be extended to thermal vapor cell experiments for broad sensing applications.
The atoms are first laser-cooled in a magneto-optical trap and then transferred to a far-detuned 1064~nm optical dipole trap (following Ref.~\cite{vendeiro2022machine}), where optical pumping is performed. The dipole-trapped atomic cloud is cooled to $14~\mu$K by gray molasses. Doppler broadening is further reduced by applying counter-propagating 780~nm probe ($1/e^2$ radius of 3.7~$\mu$m) and 479~nm control ($1/e^2$ radius of 6.8~$\mu$m) beams. The atomic cloud has a root-mean-square (RMS) radius of $10~\mu$m along the probe propagation direction and an OD of approximately 0.4.

This low OD is unfavorable for MW sensing, as spectral contrast increases with larger OD. In addition, the use of a tightly confined optical dipole trap limits the atomic sample length and probe beam waist, while providing no benefit for sensing applications. A sensing-optimized apparatus would instead employ an extended atomic medium (mm–cm scale), large probe and control beam waists, and operation at high duty cycle or in continuous mode, which could substantially improve the achievable sensitivity. Nevertheless, loading atoms into an optical dipole trap is routine in quantum optics experiments, and our demonstration illustrates the utility of this method beyond dedicated sensing platforms.

The MW fields are delivered by a broadband horn antenna (FMWAN159-10SF, Fairview Microwave) driven by a signal generator (Anritsu 68369A/NV). 
Despite using a single antenna, the MW field at the atoms contains all $\sigma^+$, $\sigma^-$ and $\pi$ polarization components due to reflections and scattering from the in-vacuo components, the vacuum chamber and nearby external structures, as is typical in cold-atom experiments~\cite{kurdak2025high}.
The DC electric field $E_{\text{DC}}$ and MW detuning $\Delta_{m}$ produce similar effects on the spectra, because both primarily shift the Rydberg S and P states differentially, making them highly correlated in the fit. Hence,
in the fitting procedure, we fix the nominal MW detuning $\Delta_m$ according to the set value of the signal generator. We verify the validity of this treatment by fitting the MW spectra immediately following DC electric field calibration~\cite{panja2024electric}.

We modulate the dipole trap with a period of 180~$\mu$s and 50\% duty cycle, and probe during the dipole-trap-off time. After 50 modulation cycles, we keep the dipole trap off to allow the atoms to disperse. 
Following a 10~ms time-of-flight, we probe for 2~ms to measure photon counts without atoms. We average these no-atom counts to obtain a normalization factor for each spectroscopy scan, which is used to normalize the transmission data. 
We then average the normalized probe transmission over multiple scans. Each spectral data point corresponds to roughly 20~ms of interrogation time. We infer the probe photon rate ($R_p$) at the apparatus from the no-atom counts, taking into account the collection and detection efficiencies of the single photon counting module (SPCM-AQRH-14-FC).

\subsection{Rydberg-mediated nonlinearity}

\begin{figure*}[ht]
    \centering
\includegraphics[width=\textwidth]{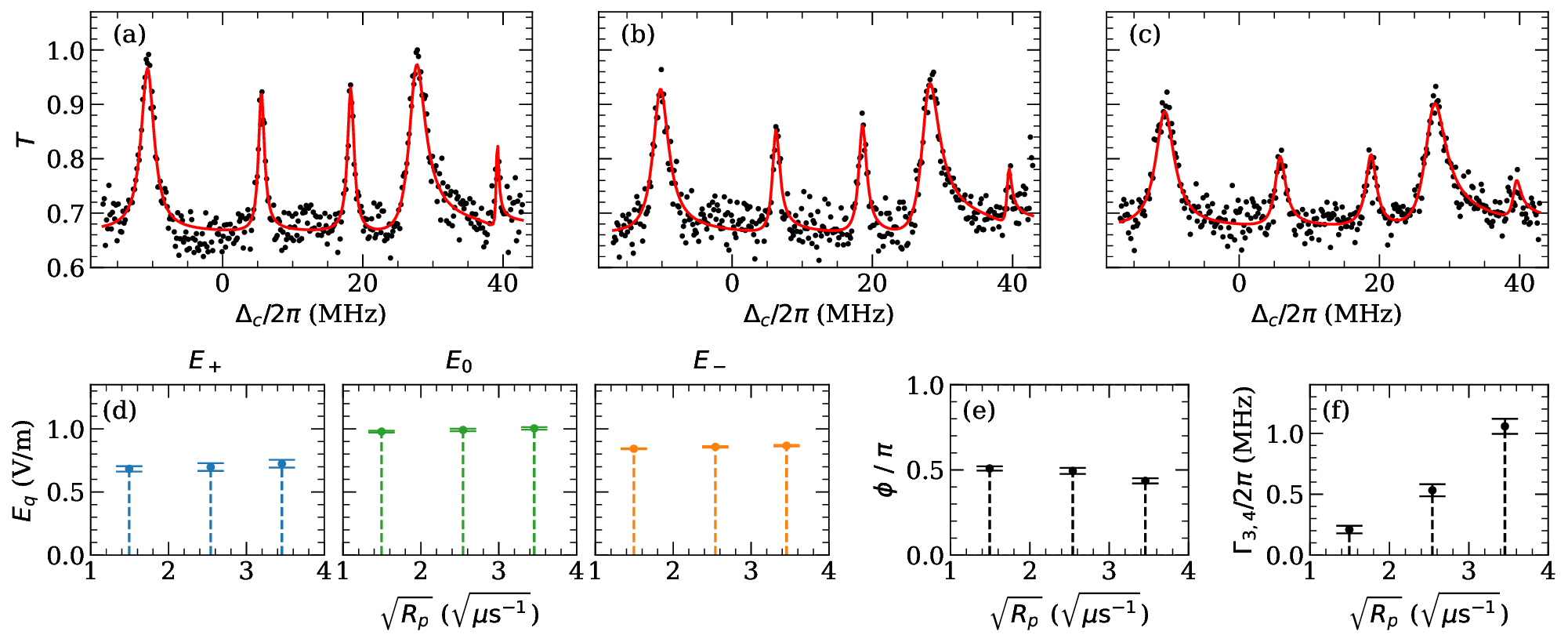}
\caption{Impact of Rydberg-mediated photon-photon interactions on parameter extraction. For all data in this figure, $\Omega_c/(2\pi)=6.65$~MHz, $B\approx8.5$~Gauss and $\Delta_m/(2\pi)=0.165$~MHz. (a-c) Probe transmission spectra for $R_p$=\{2.23, 6.46, 11.94\}~$\mu$s$^{-1}$, corresponding to estimated probe Rabi frequencies $\Omega_p/(2\pi)=\{0.11, 0.18, 0.24\}$~MHz (black dots: measured probe transmission data; red lines: fitted curves).  (d-f) Extracted MW electric-field amplitudes for $E_+$ (blue), $E_0$ (green), and $E_-$ (orange), phase $\phi=2\phi_0-\phi_+-\phi_-$, and Rydberg dephasing $\Gamma_{3,4}$ ($\Gamma_3=\Gamma_4$ constrained), respectively, as functions of the probe rate. Dashed lines serve as visual guides.
}
\label{fig: multi-level MW EIT nonlinearity}
\end{figure*}

Although $\Omega_p$ explicitly appears in Eq.~\ref{eq: transmission}, at the probe rate used in our experiments the resulting spectra are expected to exhibit virtually no dependence on $\Omega_p$. In the weak-probe limit, the steady-state solution for $\rho_{21}$ is proportional to $\Omega_p$, so this dependence cancels in the transmission spectrum. Accordingly, $\Omega_p$ is fixed in fits based on the inferred probe rate and beam waist~\cite{wang2026nonlinear}, and reasonable deviations do
not affect the extracted MW parameters. 
The nonlinearity relevant here is instead the many-body nonlinearity arising from Rydberg-mediated probe-photon interactions. Our prior work~\cite{wang2026nonlinear} showed that these interaction effects can be captured effectively by increased $\Gamma_{3}$, $\Gamma_{4}$, while still allowing accurate extraction of the MW parameters with the linear model. As a result, in our fitting procedure, the interaction-induced nonlinearity is not explicitly included in the model, but accounted by the effective dephasing rates $\Gamma_{3}$ and $\Gamma_{4}$.

Here, we verify that the Rydberg-mediated nonlinearity manifests primarily as additional dephasing by measuring spectra at increasing probe rates. Fig.~\ref{fig: multi-level MW EIT nonlinearity}(a) is close to the linear regime, where further reducing the probe rate would produce at most a small change in the spectra. Fig.~\ref{fig: multi-level MW EIT nonlinearity}(b) and (c), acquired at probe rates 2.9 and 5.3 times that of (a), respectively, are in a moderate nonlinear regime, where the interaction-induced dephasing is smaller than the EIT linewidth, i.e. $\Gamma_3+\Gamma_4<\Omega_c^2/\Gamma_2$. As the probe rate increases, the fitted MW field amplitudes and phase $\phi \equiv 2\phi_0-\phi_+-\phi_-$ remain nearly unchanged, showing no systematic trend, whereas $\Gamma_{3}$ and $\Gamma_{4}$ increases substantially (Fig.~\ref{fig: multi-level MW EIT nonlinearity}(d-f)). These results show that operating in the moderate nonlinear regime does not introduce systematic errors in the extracted MW parameters. Although a higher probe rate generally reduces photon shot noise, the resulting improvement is partly offset by saturation, so sensitivity is optimized at an intermediate probe rate. For the remainder of this manuscript, we therefore work in a regime reasonably close to this optimum.
In addition, constraining $\Gamma_3=\Gamma_4 (\equiv \Gamma_{3,4})$ to reduce the number of fit parameters, or allowing them to vary independently to slightly improve the least-squares residual in some datasets, does not statistically affect the extracted MW parameters.

\subsection{Phase identifiability}

\begin{figure}[ht]
\centering
\includegraphics[width=0.85\columnwidth]{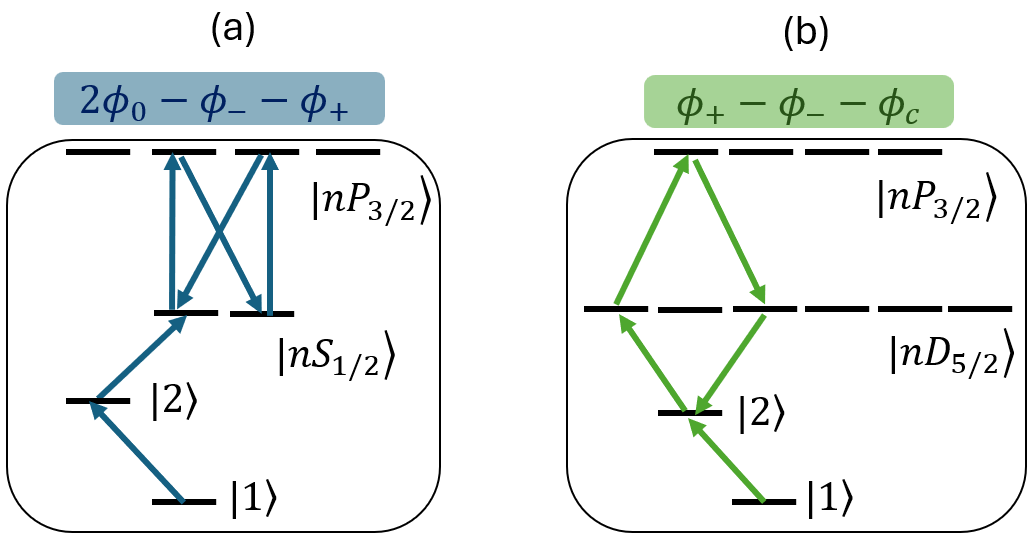 }
\caption{Internal-state Rydberg atom interferometer. (a) $\ket{nS_{1/2}}\leftrightarrow\ket{nP_{3/2}}$ transition in the presence of a $\sigma^+$-polarized control beam. (b) $\ket{nD_{5/2}}\leftrightarrow\ket{nP_{3/2}}$ transition in the presence of a linearly-polarized control beam, which is a superposition of $\sigma^+$ and $\sigma^-$. In this configuration, both types of interferometric loops are present: the MW-only loop with phase $2\phi_0-\phi_+-\phi_-$ and the control-referenced loop with phase $\phi_+-\phi_- - \phi_c$.  }
\label{fig:angle_vs_phase}
\end{figure}

In the present implementation, the detection is only sensitive to one independent MW phase combination. This limitation arises because the system has rotational symmetry about the quantization axis. For instance, $\phi_+-\phi_-$ specifies the orientation of the MW polarization ellipse in the transverse plane. Without a reference direction in that plane, rotating the transverse basis by angle $\alpha$ shifts $\phi_q\rightarrow \phi_q+q\alpha$, so any loop phase $\sum_qc_q\phi_q$ changes by $(\sum_qc_qq)\alpha$, where $c_q$ is the net number of absorption events minus emission events for a polarization $q$ photon in the loop. $\sum_qc_qq$ is the net change of the angular momentum projection, and constraint $\sum_qc_qq=0$ on interferometric loops is the mechanism by which the rotational symmetry is enforced. Fig.~\ref{fig:angle_vs_phase}(a) shows the one available loop in the current measurement scheme, which allows one phase combination to be extracted, while the other is undetermined due to this symmetry.

Providing a transverse reference removes the phase ambiguity, for example, by involving another field's relative phase between its polarization components or by performing spectroscopic measurements along multiple quantization axes. Specifically, Fig.~\ref{fig:angle_vs_phase}(b) illustrates one such approach: by using a linearly polarized control field and the $\ket{D_{5/2}}\leftrightarrow\ket{P_{3/2}}$ transition, another interferometric loop becomes available whose phase necessarily involves the relative phase between the $\sigma^+$ and $\sigma^-$ polarization components of the control field ($\phi_c$). Assigning a value to this relative control phase establishes the transverse reference. We leave the experimental study of complete phase reconstruction to future work (see Supplement 1 for further discussions). Here, we focus on measurements using the current scheme.

By varying the rotation angle of the horn while keeping it pointed at the atoms, we observe gradual variations in the fitted phase $\phi$. However, because reflected and scattered fields contribute to the local MW field at the atoms, rotating the horn changes not only $\phi$ but also the field-component amplitudes $E_q$. 
Apart from the trivial $2\pi$ periodicity, the present scheme does not distinguish between $\phi$ and $-\phi$. This sign ambiguity arises because the phase-sensitive contribution contains the closed-loop coupling product together with its Hermitian-conjugate reverse-loop product. Their sum gives a dependence proportional to $e^{i\phi}+e^{-i\phi}=2\cos\phi$~\cite{arimondo2018laser,xue2004phase}, a form typical of interferometric measurements. Accordingly, we report $\phi$ only in the interval $[0,\pi]$.

\subsection{3D microwave measurement}
\begin{figure*}[ht]
\centering
\includegraphics[width=2\columnwidth]{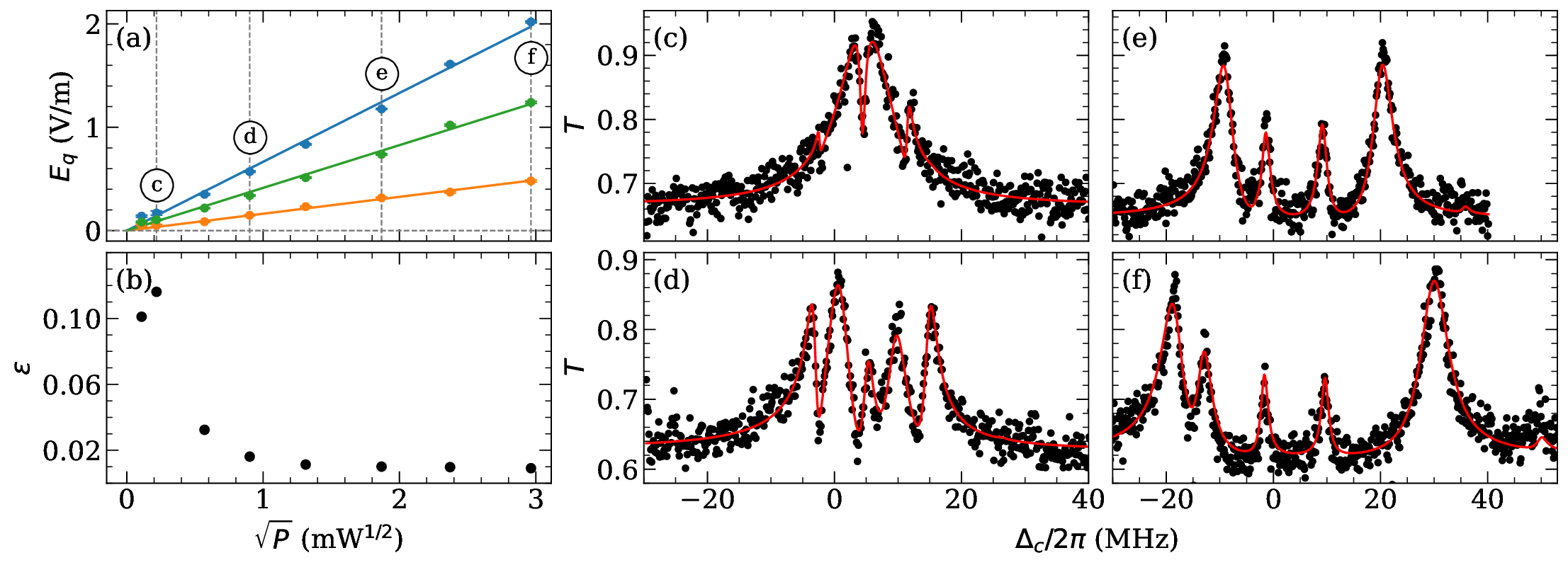}
\caption{Microwave measurement results as functions of the source power $P$ and example spectra. For all data in this figure, $\Omega_c/(2\pi)\approx7.6$~MHz, $B\approx3.9$~G, $R_p\approx10.7~\mu$s$^{-1}$ and $\Delta_m/(2\pi)=0.165$~MHz. The MW horn is positioned 20~cm from the atomic position, and no stub tuner is used. (a) MW electric-field amplitudes of each polarization $E_q$ versus $\sqrt{P}$. The solid dots are fitted values for $E_+$ (blue), $E_0$ (green), and $E_-$ (orange). The solid lines are linear fits to the solid dots with , constrained to zero intercept. Dashed lines serve as visual guides, marking zero field and the MW powers corresponding to the spectra in (c–f). The 5th dataset from the left corresponds to the spectrum in Fig.~\ref{fig:setup & levels & spectrum}(c). (b) Geometric mean of the relative uncertainties of the three polarization amplitudes, $\varepsilon$. (c–f) Representative spectra at different MW powers (black dots: measured probe transmission data; red lines: fitted curves).   }
\label{fig:power and frequency dependence}
\end{figure*}

As shown in Fig.~\ref{fig:power and frequency dependence}(a), each electric-field polarization amplitude $E_q$ depends linearly on $\sqrt{P}$, where $P$ is the nominal power output of the MW signal generator. At the two lowest powers, the spectra contain only three clearly resolved spectral peaks, leaving the phase $\phi$ underdetermined and producing relatively large uncertainties in the field amplitudes.
To provide an overall measure of uncertainty in the inferred MW-field amplitudes, we define
$\varepsilon \equiv \left(\prod_q \frac{\sigma_q}{E_q}\right)^{1/3}$,
which is the geometric mean of the relative uncertainties of the three polarization amplitudes, with $\sigma_q$ denoting the fit uncertainty of $E_q$. Fig.~\ref{fig:power and frequency dependence}(f) shows $\varepsilon$ for the fits in panel (a), using the standard deviations obtained from the covariance matrix of the nonlinear least-squares fit.

Such uncertainty estimates are only strictly valid when the model is approximately linear in all fit parameters near the optimum. Due to the large number of parameters in our model, these conditions may not always be satisfied. We thus perform bootstrap analysis~\cite{davison1997bootstrap,hesterberg2011bootstrap} on representative datasets to assess the robustness of these uncertainty estimates. In both nonparametric case-resampling bootstrap and parametric bootstrap (procedures and detailed results described in Supplement 1), the medians of the bootstrap replicas agree with the fitted values, all lying well within the single-fit standard deviations and 68\% confidence intervals. Furthermore, the bootstrap intervals do not reveal a systematic bias in the uncertainty estimates from the single nonlinear least-squares fit, although individual parameter uncertainties can differ by up to about a factor of three.

The exception to this behavior occurs when the fitted phase $\phi$ is near the boundaries, i.e. 0 or $\pi$. In this scenario, bootstrap analyses result in highly non-(half)Gaussian distributions while single-fit standard deviations cover nearly the entire $\pi$ range. We find the same behavior even when no boundary is imposed on the fitted phase, or when the fit is reparameterized to avoid periodicity.
We also examined the parameter dependence and did not find a qualitatively different landscape near the phase boundaries, such as an anomalously flat or sharp dependence. In the power-dependence data shown here, the fitted phases that are well determined are all close to $\pi$. Because the boundary-sensitive phase uncertainties do not provide an additional useful comparison across powers, we focus this figure on the MW-field amplitudes.

Fig.~\ref{fig:power and frequency dependence}(c-f) show representative spectra and fits at different MW powers. 
The precision of the extracted MW parameters depends on the details in the spectral structure. Although spectra with three or fewer resolved peaks generally do not contain enough information to well determine all MW parameters, simply resolving more peaks does not necessarily guarantee smaller uncertainties. At fixed control-parameter values, increasing the MW power also does not always increase the number of resolved spectral peaks. The choice of control parameters can affect how well the MW parameters are constrained. $E_{\text{DC}}$ and $B$ are two parameters that can be tuned in situ and adjusted based on the MW fields to optimize sensing capability. Here, we do not assume prior knowledge of the sensing parameters and do not intentionally use $E_{\rm DC}$ as a control parameter to improve parameter sensitivity. Rather, we cancel the bulk of the ambient electric field but do not maintain the cancellation. As a result, our $E_{\text{DC}}$ represents a residual electric field arising from daily drifts.

\begin{figure*}[ht]
\centering
\includegraphics[width=0.75\textwidth]{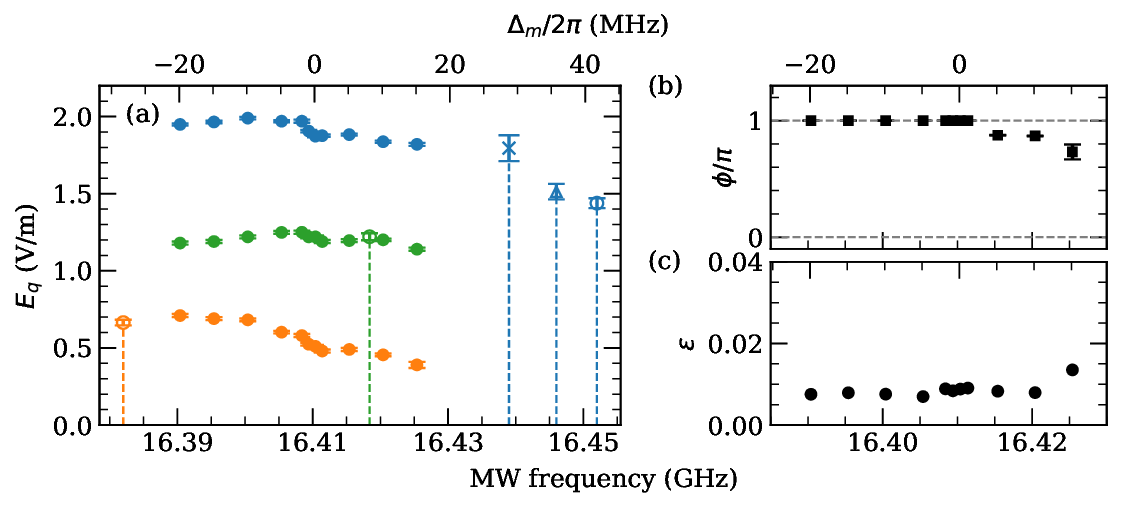}
\caption{Microwave measurement results as functions of frequency. The MW horn is positioned 20~cm from the atomic position, and no stub tuner is used. (a) Extracted MW electric-field amplitudes of each polarization versus MW detuning for $E_+$ (blue), $E_0$ (green), and $E_-$ (orange). The solid dots are fitted values from multi-level spectra. For all multi-level spectra in this figure, $\Omega_c/(2\pi)\approx9.0$~MHz, $B=3.8$~G, $R_p\approx8.9~\mu$s$^{-1}$ and $P=9.44$~dBm. The open circles are fitted values multiplied by 5.2 from four-level spectra at $B=18.5$~G and $P=-4.88$~dBm. The triangle (cross) marks the fitted value of $E_+$ multiplied by 13.6 (27.3) at $B=15.7$~G (13.0~G) and $P=-13.23$~dBm ($-19.28$~dBm). Except for the magnetic field $B$ and source power $P$, all spectra in this figure were taken under similar experimental conditions. Vertical dashed lines serve as visual guides for the four-level results. (b,c) Extracted phase $\phi/\pi$ and mean relative uncertainty $\varepsilon$. The dashed lines in (b) mark the phase boundaries.  }
\label{fig:freq_vs_E_no_tuner}
\end{figure*}

Other control parameters can be substantially improved through apparatus design. For an optically thin medium like ours, a larger OD is better, because it lowers the off-resonant transmission baseline. Reducing $\Gamma_{3,4}$ increases the EIT peak height, and both effects enhance the spectral contrast. If the probe beam waist is larger, then for the same probe rate, the interaction-induced dephasing will be reduced. A larger control beam waist will reduce the spatial inhomogeneity, which we attribute as dominating the broadening mechanism at larger $\Omega_c$. We therefore choose an $\Omega_c$ that the EIT linewidth $\Omega_c^2/\Gamma_2$ exceeds the effective decoherence, while avoiding a regime where control-field inhomogeneity dominates the effective $\Gamma_{3,4}$.

\begin{figure*}[ht]
\centering
\includegraphics[width=\textwidth]{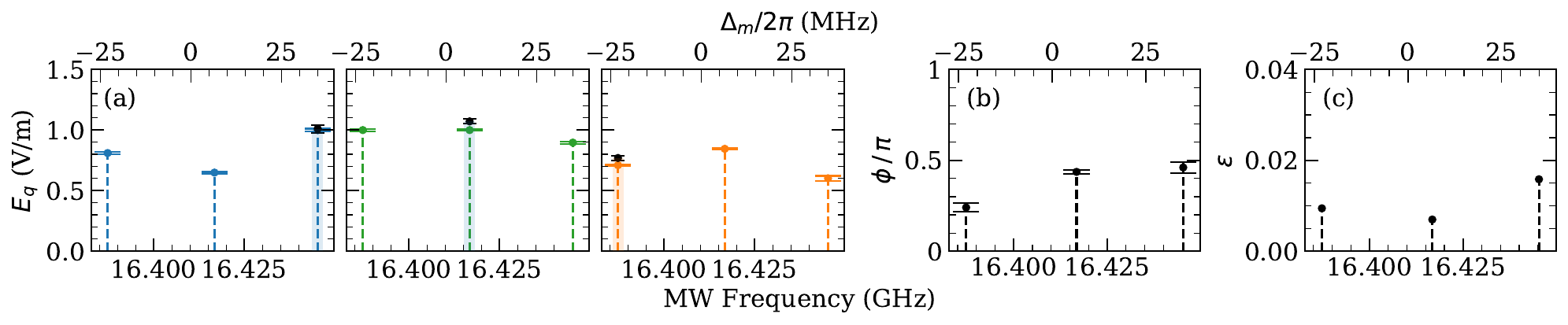}
\caption{Comparison between three multi-level measurements and the corresponding four-level measurements at matched MW frequencies. The MW horn is positioned 40~cm from the atomic position, using the stub tuner to minimize reflections. For all data in this figure, $\Omega_c/(2\pi)\approx6.9$~MHz and $R_p=6.51~\mu$s$^{-1}$. From high to low MW frequency, $B=\{11.1,6.1,3.7\}$~G. (a-c) Extracted MW electric-field amplitudes of each polarization versus MW detuning for $E_+$ (blue), $E_0$ (green), and $E_-$ (orange). The black dots are four-level fit results for $E_+$ (a), $E_0$ (b) and $E_-$ (c) with $B=\{15.4,15.8,15.7\}$~G, respectively. (d,e) Extracted phase $\phi/\pi$ and mean relative uncertainty $\varepsilon$. Vertical dashed lines serve as visual guides for the multi-level results. }
\label{fig:example spectrum and MW frequency & power}
\end{figure*}

Likewise, the frequency span over which the spectroscopy can efficiently resolve the MW fields is also tied to the detectable spectral features. Fig.~\ref{fig:freq_vs_E_no_tuner} shows the extracted MW fields across a frequency span of approximately 40~MHz. We observe variations in each polarization amplitude. From $-20$ to $2$~MHz, the fitted phase values remain pinned at $\pi$ to four decimal places, whereas above 2~MHz, phase $\phi$ deviates smoothly away from $\pi$. The mean relative uncertainty $\varepsilon$ remains low at 1\% level. 

To both cross-check the fit results and show the distinction from the  traditional Autler-Townes (AT) splitting approach, we apply a large magnetic field $B=18.5~$G and sufficiently weak MW fields tuned into resonance with one of the shifted $\sigma^+$, $\pi$, or $\sigma^-$ transitions, such that only one of the three Zeeman sublevels $\ket{4_{1/2}}$, $\ket{4_{-1/2}}$, and $\ket{4_{-3/2}}$ is coupled to $\ket{3_{-1/2}}$. This isolates an effective four-level system. In this system, we can determine each ${E}_q$ at its corresponding resonant MW frequency.
The extracted values of the three polarization components are shown as open circles in Fig.~\ref{fig:freq_vs_E_no_tuner}(a). The three effective four-level measurements are performed under the same conditions except for the MW frequency, which is tuned to resonance with the corresponding polarization-selective transition. Relative to the multi-level measurements, these four-level measurements use a different magnetic field and a MW source power $P$ lower by 14.32~dB, corresponding to a factor of 5.2 in electric-field amplitude. For comparison with the multi-level results, the electric fields extracted from the effective four-level spectra are multiplied by 5.2 in the plot. 

We extract each MW polarization component using both Gaussian peak fits, corresponding to the AT-splitting analysis, and effective four-level model fits. The measured AT splittings are statistically consistent with the generalized MW Rabi frequency from the four-level model. Because our data have $\lesssim1~$MHz MW detuning from the transition frequency shifted by the DC electric and magnetic fields, we show the four-level fit results in the figures.

Because the multi-level and effective four-level measurements of $E_+$ are separated by a relatively large MW-frequency difference, we measure two additional $E_+$ points at smaller $\Delta_m$ using the effective four-level system. For these measurements, we lower the MW power further so that the off-resonant couplings omitted from the effective four-level model remain negligible compared with the statistical uncertainties. The resulting two effective four-level measurements are shown by the  triangle and cross markers. They correspond to MW-power reductions of $22.67$ and $28.72$~dB relative to the multi-level spectra, respectively, and are scaled by the corresponding factors in the plot. 

All effective four-level fit results are consistent with the trend projected from our multi-level results, but they cannot be used to reconstruct the 3D MW electric fields because they are measured at different frequencies, and the frequency dependence is evident. Under circumstances like ours, our multi-level approach becomes particularly useful. To further test the frequency dependence, we use a stub tuner to minimize reflection at each frequency (see Supplement 1), while data in Figs.~\ref{fig:power and frequency dependence} and \ref{fig:freq_vs_E_no_tuner} are taken without the stub tuner in the circuit. The horn is also repositioned to obtain different phase values. More importantly, we empirically adjust the bias magnetic field $B$ values to extend the frequency coverage, allowing the multi-level and four-level spectra to be taken at exactly the same MW frequencies.

Again, we observe variations in the amplitudes $E_q$ and phase $\phi$ over an approximately 60~MHz span (Fig.~\ref{fig:example spectrum and MW frequency & power}). 
Our MW wavelength is 1.8~cm. Structures with length scales comparable to or larger than this can make the microwave field spatially complex and speckle-like~\cite{kaina2014shaping}. Examples include in-vacuo metals such as electrodes and dispensers, as well as metals surrounding the glass cell, including coils and their mounts, vacuum hardware, optical breadboards and optical mounts. In this environment, the local MW field sampled by the atoms is a coherent sum of the directly incident field and fields that are scattered or reflected by these surrounding structures.
We hypothesize that interference among these propagation paths is responsible for the observed frequency dependence, since small changes in frequency can change their relative phases and thereby alter the local field distribution sampled by the atoms. For a simple argument, each scattered or reflected wave accumulates an additional phase $ L \,\delta \Delta_m / c$ when the frequency
changes by $\delta\Delta_m$, where $L$ is its path length to the atoms. For a horn-to-atom path length $L\approx 40~$cm, a frequency change
of $\delta\Delta_m/(2\pi) = 60$~MHz shifts the phase by approximately $0.2\pi$.

\section{Conclusion and outlook}
In conclusion, we have demonstrated a self-calibrated method for extracting 3D MW electric-field information from Rydberg EIT spectra in a cold-atom apparatus not specifically designed for microwave sensing. By fitting the Zeeman-resolved spectral structure to a multi-level model, we simultaneously recover the amplitudes of the $\sigma^+$, $\pi$, and $\sigma^-$-polarized components, together with one identifiable relative phase combination. This method extracts all measurable MW field parameters from a spectrum acquired at a single MW frequency, which is particularly valuable when changes in frequency or other settings alter the field. We further showed that the extracted microwave parameters remain robust in the presence of moderate Rydberg-mediated nonlinearity. Bootstrap analysis supports the fit-derived uncertainty estimates despite the large number of parameters and the associated complex fitting landscape. These results establish multi-level Rydberg spectroscopy as a practical and broadly applicable tool for MW field characterization.

In the future, we plan to use the $\ket{D_{5/2}}\leftrightarrow\ket{P_{3/2}}$ transition to fully extract the 3D vector MW field, including both independent phase combinations. Our method represents a different perspective on the role of Zeeman sublevels: here they constitute a resource for extracting additional information, whereas in conventional AT spectroscopy they are a source of spectral complexity~\cite{PhysRevA.109.L021702,schlossberger2026MW} that needs to be understood for precise measurements. Our detailed study of Zeeman-resolved spectroscopy may in turn inform broader developments in atomic sensing.

We will also develop an automatic, adaptive procedure to optimize control parameters in real time, assisted by machine-learning-based surrogate models that reduce computational cost. This will enable systematic diagnosis of how spectral features constrain fitted parameters and clarify the origin of anomalous phase uncertainty near boundaries. Furthermore, the resulting framework may shed light on information extraction and adaptive protocol design in broader contexts.

\noindent\textbf{Funding.} AFOSR FA9550-22-1-0327 and FA9550261B046.

\noindent\textbf{Acknowledgment.} We gratefully acknowledge Chen-Lung Hung, Deniz Kurdak, J. V. Porto, S. L. Rolston, Matthew T. Eiles and Miaoyuan Liu for valuable discussions. 

\noindent\textbf{Disclosures.} The authors declare no conflicts of interest.

\noindent\textbf{Data availability.}
The data that support the findings of this study are openly available in Zenodo at https://doi.org/10.5281/zenodo.19741036.

\noindent\textbf{Supplemental document.} See Supplement 1 for supporting content.

\clearpage
\onecolumngrid

\begin{center}
\textbf{\large Supplementary Information:}\\
\textbf{\large Self-calibrated multiparameter measurement of three-dimensional microwave fields}
\end{center}

\setcounter{section}{0}
\setcounter{figure}{0}
\setcounter{table}{0}
\renewcommand{\thesection}{S\Roman{section}}
\renewcommand{\thefigure}{S\arabic{figure}}
\renewcommand{\thetable}{S\arabic{table}}

\section{Parameter calibration}
\label{sec:Parameter calibration}
\begin{figure}[ht]
\centering
\includegraphics[width=0.8\textwidth]{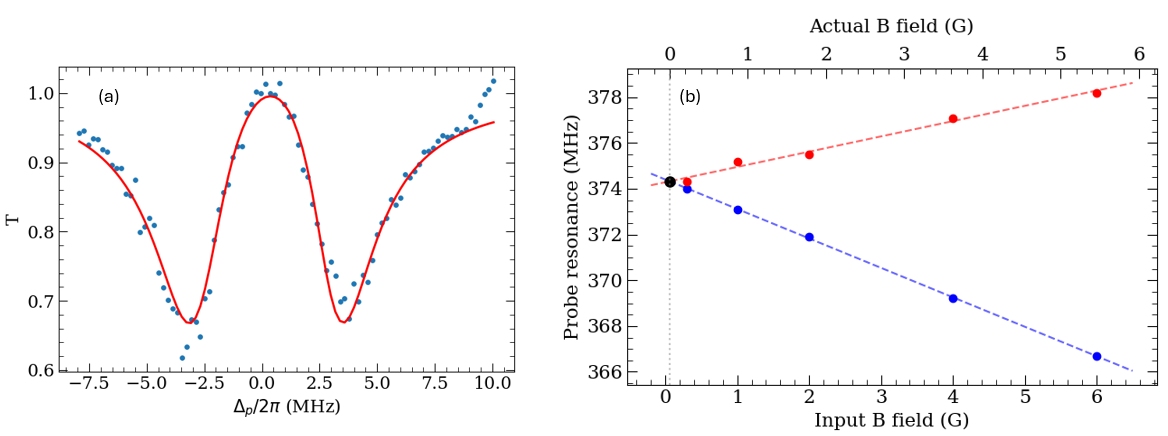}
\caption{Parameter calibrations. (a) Calibration of $\Omega_c$ by fitting (red curve) to a EIT measurement (blue dots; transmission $T$ versus probe detuning $\Delta_p/(2\pi)$). (b) B field calibration using the $\sigma^-$ probe resonance shift due to Zeeman splittings. The zero B field is determined from the crossing point (vertical dashed line) of the $\sigma^-$ (blue dots and linear fit) and $\sigma^+$ (red dots and linear fit) probe resonances. The vertical axis is the measured probe resonance with respect to the probe laser locking point.}
\label{fig:3-level fit}
\end{figure}
Many parameters, namely, $\Omega_p , \Omega_c , B, E_\text{DC}, \Gamma_{3,4} \textrm{ and OD}$ can be pre-determined. $\Omega_p$ is determined from the measured photon-rate without atoms and the $1/e^2$ radius of the probe beam ($3.7~\mu$m). It then has to be scaled down by a factor of 1.6 for better agreement with data in~\cite{wang2026nonlinear}, representing a global uncertainty likely associated with the effective probe beam size at the atomic cloud. $\Omega_c, \Gamma_{3,4}$ and OD are measured from a  probe frequency scan on a 3-level EIT scheme ($\ket{1} \rightarrow \ket{2} \rightarrow \ket{3_{-1/2}}$) without the MW (see Fig.~\ref{fig:3-level fit}(a) for an example of this fit). We find that the measured $\Omega_c$ from the $1/e^2$ radius of the control beam ($6.8~\mu$m) has a similar scaling factor of 1.4 with the $\Omega_c$ obtained from the EIT fit. Moreover, we treat OD as a fit parameter for the MW measurement in the main text as it is nearly independent from other parameters (see Fig.~\ref{fig:parametric bootstrap}(a)) and measuring it before every MW scan is time-consuming. $\Gamma_{3}$ and $\Gamma_{4}$ are also treated as fit parameters, as they capture the additional dephasing induced by Rydberg-mediated photon-photon interactions~\cite{wang2026nonlinear} that are not present in the 3-level EIT scheme. $E_{\text{DC}}$ can be independently calibrated~\cite{panja2024electric}; however, as $E_\text{DC}$ may vary by  $\approx20$~mV/cm in a day, measuring it using the method in~\cite{panja2024electric} before every MW scan is a time-consuming process. Therefore, $E_{\text{DC}}$ is treated as a fit parameter. 

$B$, the magnetic field along the quantization axis, can be obtained by measuring the shift in the probe resonance frequency on the $\ket{1}\rightarrow\ket{2}$ transition (without the control beam and MW) as the ``input'' B field is varied, as shown in Fig.~\ref{fig:3-level fit}(b). The probe resonance is presented as the difference in frequency with respect to the laser locking point. The locking scheme for the probe laser is similar to that used for the control laser and is described in our previous work (see Appendix A of Ref.~\cite{panja2024electric}). The 780-nm probe beam from the TOPTICA Photonics laser system passes through an independent fiber electro-optic modulator (PM-0S5-10-PFA-PFA-780, EOSpace) before being directed into the same optical cavity used to stabilize the control laser. The resulting error signal is fed back to the DLC pro controller through an additional FALC module to stabilize the probe laser frequency. To find the ambient B field at the nominal zero input field, we flip the polarization of the probe beam to $\sigma^+$ and do a similar scan. From linear fits and extrapolations on the two branches of the probe resonance vs input B field graph, we can find the ambient (0.06~G) and the actual B field. The actual B field is calibrated using the $\sigma^-$ linear fit, as the $\sigma^-$ line ($\ket{1} \rightarrow \ket{2}$ transition) is not affected by the optical pumping effect, if present. This B field calibration is a reference, but is not actually used in the fit (see Sec.~\ref{sec:Fit procedure}). The intersection of the two lines can also be used to determine the probe resonance frequency ($\nu_p^\text{abs}=374.32$~MHz) at zero magnetic field. This number should be fixed since probe transition is insensitive to DC E field. However, owing to systematic drifts in the laser locking scheme, this value can vary by up to approximately $1$~MHz over time. Therefore, throughout this work we explicitly indicate the data acquisition dates to provide context for the drift of $\nu_p^\text{abs}$. The blue data points in Fig.~\ref{fig:3-level fit}(b) were acquired on 2026/01/18, while the red data points were acquired on 2026/03/10.

\section{Fit procedure and results}
\label{sec:Fit procedure}
\subsection{Multi-level}

In the MW parameter extraction process described in the main text, fixed parameters are the microwave detuning $\Delta_{m}$ (determined directly from MW source), and the probe Rabi frequency $\Omega_p$ (obtained independently from the calibrated probe photon count rate). The control-laser Rabi frequency $\Omega_c$ is also fixed independently from this separate three-level EIT measurement performed on the same day using the method in Sec.~\ref{sec:Parameter calibration}. The free fitting parameters are the three microwave polarization components,$\{E_{+}, E_{0}, E_{-}\}$, the phase $\phi$, the ambient DC electric field $E_{\text{DC}}$, the optical depth ${\rm OD}$, two Rydberg state dephasing $\Gamma_4$ and $\Gamma_3$, and two probe frequency parameters $\nu_p^\text{abs}$ and $\nu_p^\text{act}$:

\[
\{E_{+},E_{-},E_{0},\phi,E_{\rm DC},{\rm OD},\Gamma_4,\Gamma_3,\nu_p^\text{abs},\nu_p^\text{act}\}.
\]

To fit the measured spectra data in the model described in the main text, we need to convert the probe and control frequency settings (frequency difference with respect to the locking point) into $\Delta_p$ and $\Delta_c$, which are defined as the detunings in the absence of magnetic field and DC electric field. Therefore, we want to know both fields when we take each spectrum data and the corresponding probe and control resonances. As mentioned in Sec.~\ref{sec:Parameter calibration}, we fit DC electric field and calibrate B field. We also found a sub-megahertz drift of probe resonance. This could result from a half-gauss level ambient magnetic field drift and an unstable probe AOM driving frequency and they have different time scales. Thus, instead of directly getting the daily B field from calibration and, to separate the two drift effects, we introduce the actual probe resonance $\nu_p^\text{act}$ of each spectrum's B field, and the absolute probe resonance $\nu_p^\text{abs}$ in the absence of B field. We fit these two together with other parameters for each spectrum. We are then able to get the B field and $\Delta_p$ of each spectrum as:
\begin{equation}
    B=\frac{\nu_p^\text{act}-\nu_p^\text{abs}}{\eta_B},
    \label{B_field}
\end{equation}
\begin{equation}
    \Delta_p = 2\pi(\nu_p^\text{park}-\nu_p^\text{abs}),
    \label{delta_p}
\end{equation}
where $\eta_B=(0.93\times3-0.70\times2)~{\rm MHz/G}=1.39~{\rm MHz/G}$, and $\nu_p^\text{park}$ is the probe frequency settings of each data point in the spectrum.
\\
The control-laser detuning is defined analogously as $\Delta_c = 2\pi(2\nu_c^\text{park}-2\nu_c^\text{abs})$ while the additional factor of $2$ arises because the control laser is generated through second-harmonic generation after frequency locking. With the existance of B field and DC E field, $\nu_c^\text{abs}$ can be calculated from 
\begin{equation}
    2\nu_c^\text{abs}=2\nu_c^\text{act}-(-\tfrac{1}{2}\alpha_{3,-1/2} E_{\rm DC}^2)-(Z_\text{3,-1/2}^\text{(2)}(B)+Z_{\rm 3,-1/2}(B)-Z_{\rm 2}(B))
    \label{nuc_abs},
\end{equation}
where $\nu_c^\text{act}$ is the control resonance under each spectrum's own B field and DC E field, and can be obtained from $\nu_p^\text{act}$ and two-photon resonance $\nu_\text{2-photon}^\text{act}$; $-\tfrac{1}{2}\alpha_{3,-1/2} E_{\rm DC}^2$ is the DC Stark shift of $\ket{3_{-1/2}}$ state; $Z_{\rm i, m_j}^{(k)}(B)$ terms are the significant enough Zeeman shifts. Substituting Eq.~(\ref{nuc_abs}) into the definition of $\Delta_c$ yields
\begin{equation}
    \Delta_c=2\pi\left(2\nu_c^\text{park}-2\nu_c^\text{act}\right)-\tfrac{1}{2}\alpha_{3,-1/2} E_{\rm DC}^2+Z_\text{3,-1/2}^\text{(2)}(B)+Z_{\rm 3,-1/2}(B)-Z_{\rm 2}(B).
    \label{delta_c}
\end{equation}
\\
We daily calibrated a two-photon resonance $\nu_\text{2-photon}^\text{act}$ using 3-level EIT ($\ket{1} \rightarrow \ket{2} \rightarrow \ket{3_{-1/2}}$), since it is not affected by B field drifts across spectra (the control Rabi frequency $\Omega_c$ is calibrated from the 3-level EIT as well since it does not change within the day). We used two equivalent methods to get $\nu_\text{2-photon}^\text{act}$: 
\\
(i) If the probe laser is fixed at $\nu_p^\text{set}$ while the control frequency is scanned, the resulting three-level EIT spectrum is fitted with a Gaussian profile. The fitted peak center of the control scan, $\nu_c^\text{fit}$, is taken as the control resonance frequency. The two-photon resonance is then
\begin{equation}
    \nu_\text{2-photon}^\text{act}=\nu_p^\text{set}+2\nu_c^\text{fit},
    \label{two_photon_resonance_park_probe}
\end{equation}
(ii) If the control laser is fixed at $\nu_c^\text{set}$ while the probe frequency is scanned, the measured three-level EIT probe spectrum is fitted using the three-level analytical equation to extract the probe resonance frequency $\nu_p^\text{fit}$ and the control detuning $\delta_c$. The two-photon resonance is then (there are alternative ways but not presented here)
\begin{equation}
    \nu_\text{2-photon}^\text{act}=\nu_p^\text{fit}+2\nu_c^\text{set}-\delta_c,
    \label{two_photon_resonance_park_control}
\end{equation}
Then for each spectrum, the control resonance can be obtained as
\begin{equation}
    \nu_c^\text{act}=\frac{\nu_\text{2-photon}^\text{act}-\nu_p^\text{act}}{2}.
    \label{control_freq_when_no_E}
\end{equation}
In practice, the fitting is performed in two steps. In the first round, $\nu_p^\text{abs}$ is allowed to vary. We found that $\nu_p^\text{abs}$ was stable over measurements taken within approximately three days, while it could drift noticeably (more than 0.4~MHz) over longer time scales, such as one week or more. Therefore, in the second round of fitting, $\nu_p^\text{abs}$ is fixed to the average value obtained from the first-round fit for data taken within the same short time window.

Here we present the fitted spectra and corresponding data points discussed in the main text. Since the first- and second-round fitting results are nearly identical, we only show the second-round fits. The fitted values of $\nu_p^\text{abs}$ obained from the first round are reported in the format “fitted/\textbf{fixed}”.

\begin{table}[htbp]
    \footnotesize
    \begin{tabular}{|c|c|c|c|c|c|c|c|c|c|c|c|c|c|c|}
        \hline
        \shortstack{$E_{+}$ \\ (V/m)} & \shortstack{$E_{-}$ \\ (V/m)} & \shortstack{$E_{0}$ \\ (V/m)} & $\phi/\pi$ & \shortstack{$E_{\textrm{DC}}$ \\ (V/m)} & OD & \shortstack{$\Gamma_{4}/2\pi$ \\ (MHz)} & \shortstack{$\Gamma_{3}/2\pi$ \\ (MHz)} & \shortstack{$\nu_p^\text{abs}$ \\ (MHz)} & \shortstack{$\nu_p^\text{act}$ \\ (MHz)} & \shortstack{$B_{\text{calc}}$ \\ (gauss)}  & \shortstack{$\bm{\Omega_{c}/2\pi}$ \\ \textbf{(MHz)}} & \shortstack{$\bm{\Omega_{p}/2\pi}$ \\ \textbf{(MHz)}} & \shortstack{$\bm{\nu_p^\textbf{park}}$ \\ \textbf{(MHz)}} & \shortstack{$\bm{\Delta_m/2\pi}$ \\ \textbf{(MHz)}} \\ 
        \hline
        0.68 & 0.84 & 0.98 & 0.51 & 0.063 & 0.403 & 0.21 & 0.21 & 374.14/\textbf{374.24} & 362.26 & 8.56 & \textbf{6.65} & \textbf{0.105} & \textbf{362.0} & \textbf{0}\\
        \hline
        0.70 & 0.857 & 0.99 & 0.49 & 0.051 & 0.409 & 0.53 & 0.53 & 374.24/\textbf{374.24} & 362.48 & 8.40 & \textbf{6.65} & \textbf{0.179} & \textbf{362.0} & \textbf{0}\\
        \hline
        0.72 & 0.867 & 1.00 & 0.44 & 0.054 & 0.395 & 1.06 & 1.06 & 373.97/\textbf{374.24} & 362.36 & 8.49 & \textbf{6.65} & \textbf{0.243} & \textbf{362.0} & \textbf{0}\\
        \hline

    \end{tabular}
    \caption{Fitted and fixed parameters (from top to bottom) corresponding to the three spectra shown in Fig.~2(a–c). $\Gamma_3$ and $\Gamma_4$ are constrained to be equal.}
    \label{table : Fit Parameters fig2}
\end{table}

\begin{table}[htbp]
    \footnotesize
    \begin{tabular}{|c|c|c|c|c|c|c|c|c|c|c|c|c|c|c|}
        \hline
        \shortstack{$E_{+}$ \\ (V/m)} & \shortstack{$E_{-}$ \\ (V/m)} & \shortstack{$E_{0}$ \\ (V/m)} & $\phi/\pi$ & \shortstack{$E_{\textrm{DC}}$ \\ (V/m)} & OD & \shortstack{$\Gamma_{4}/2\pi$ \\ (MHz)} & \shortstack{$\Gamma_{3}/2\pi$ \\ (MHz)} & \shortstack{$\nu_p^\text{abs}$ \\ (MHz)} & \shortstack{$\nu_p^\text{act}$ \\ (MHz)} & \shortstack{$B_{\text{calc}}$ \\ (gauss)}  & \shortstack{$\bm{\Omega_{c}/2\pi}$ \\ \textbf{(MHz)}} & \shortstack{$\bm{\Omega_{p}/2\pi}$ \\ \textbf{(MHz)}} & \shortstack{$\bm{\nu_p^\textbf{park}}$ \\ \textbf{(MHz)}}  & \shortstack{$\bm{\Delta_m/2\pi}$ \\ \textbf{(MHz)}}\\ 
        \hline
        0.14 & 0.05 & 0.083 & X & 0.0555 & 0.436 & 0.7 & 2.8 & 375.00/\textbf{374.80} & 369.28 & 3.95 & \textbf{7.6} & \textbf{0.230} & \textbf{369.4} & \textbf{0.135}\\
        \hline
        0.17 & 0.05 & 0.105 & X & 0.0592 & 0.409 & 0.42 & 2.1 & 374.86/\textbf{374.80} & 369.39 & 3.87 & \textbf{7.6} & \textbf{0.229} & \textbf{369.4} & \textbf{0.135}\\
        \hline
        \color{blue}0.351 & \color{blue}0.086 & \color{blue}0.218 & \color{blue}1.0 & \color{blue}0.0617 & \color{blue}0.412 & \color{blue}0.5 & \color{blue}2.4 & \color{blue}374.66/\textbf{374.80} & \color{blue}369.33 & \color{blue}3.91 & \color{blue}\textbf{7.3} & \color{blue}\textbf{0.226} & \color{blue}\textbf{369.4} & \color{blue}\textbf{0.135}\\
        \hline
        \color{brown}0.571 & \color{brown}0.148 & \color{brown}0.339 & \color{brown}1.0 & \color{brown}0.0576 & \color{brown}0.468 & \color{brown}0.44 & \color{brown}3.3 & \color{brown}374.57/\textbf{374.67} & \color{brown}369.30 & \color{brown}3.84 & \color{brown}\textbf{7.3} & \color{brown}\textbf{0.233} & \color{brown}\textbf{369.4} & \color{brown}\textbf{0.135}\\
        \hline
        0.835 & 0.232 & 0.513 & 1.0 & 0.0659 & 0.442 & 0.5 & 4.0 & 374.65/\textbf{374.80} & 369.36 & 3.89 & \textbf{7.6} & \textbf{0.230} & \textbf{369.4} & \textbf{0.135}\\
        \hline
        1.177 & 0.317 & 0.738 & 1.0 & 0.0638 & 0.440 & 1.0 & 1.9 & 374.81/\textbf{374.80} & 369.34 & 3.90 & \textbf{7.6} & \textbf{0.230} & \textbf{369.4} & \textbf{0.135}\\
        \hline
        \color{purple}1.610 & \color{purple}0.373 & \color{purple}1.021 & \color{purple}1.0 & \color{purple}0.0438 & \color{purple}0.492 & \color{purple}1.0 & \color{purple}3.3 & \color{purple}374.79/\textbf{374.80} & \color{purple}369.43 & \color{purple}3.84 & \color{purple}\textbf{8.2} & \color{purple}\textbf{0.231} & \color{purple}\textbf{369.4} & \color{purple}\textbf{0.135}\\
        \hline
        \color{purple}2.018 & \color{purple}0.48 & \color{purple}1.24 & \color{purple}0.92 & \color{purple}0.042 & \color{purple}0.482 & \color{purple}0.9 & \color{purple}3.4 & \color{purple}374.83/\textbf{374.80} & \color{purple}369.32 & \color{purple}3.92 & \color{purple}\textbf{8.2} & \color{purple}\textbf{0.232} & \color{purple}\textbf{369.4} & \color{purple}\textbf{0.135}\\
        \hline

    \end{tabular}
    \caption{Fitted and fixed parameters (from top to bottom) corresponding to each data point (from left to right) in Fig.~4(a) .}
    \label{table : Fit Parameters fig4}
\end{table}

\begin{table}[htbp]
    \footnotesize
    \begin{tabular}{|c|c|c|c|c|c|c|c|c|c|c|c|c|c|c|}
        \hline
        \shortstack{$E_{+}$ \\ (V/m)} & \shortstack{$E_{-}$ \\ (V/m)} & \shortstack{$E_{0}$ \\ (V/m)} & $\phi/\pi$ & \shortstack{$E_{\textrm{DC}}$ \\ (V/m)} & OD & \shortstack{$\Gamma_{4}/2\pi$ \\ (MHz)} & \shortstack{$\Gamma_{3}/2\pi$ \\ (MHz)} & \shortstack{$\nu_p^\text{abs}$ \\ (MHz)} & \shortstack{$\nu_p^\text{act}$ \\ (MHz)} & \shortstack{$B_{\text{calc}}$ \\ (gauss)}  & \shortstack{$\bm{\Omega_{c}/2\pi}$ \\ \textbf{(MHz)}} & \shortstack{$\bm{\Omega_{p}/2\pi}$ \\ \textbf{(MHz)}} & \shortstack{$\bm{\nu_p^\textbf{park}}$ \\ \textbf{(MHz)}}  & \shortstack{$\bm{\Delta_m/2\pi}$ \\ \textbf{(MHz)}}\\ 
        \hline
        1.949 & 0.71 & 1.18 & 1.0 & 0.034 & 0.297 & 0.62 & 1.3 & 374.66/\textbf{374.67} & 369.37 & 3.79 & \textbf{10.0} & \textbf{0.207} & \textbf{369.4} & \textbf{-19.865}\\
        \hline
        1.965 & 0.69 & 1.19 & 1.0 & 0.039 & 0.298 & 0.72 & 1.6 & 374.70/\textbf{374.67} & 369.30 & 3.84 & \textbf{10.0} & \textbf{0.206} & \textbf{369.4} & \textbf{-14.865}\\
        \hline
        1.990 & 0.682 & 1.22 & 1.0 & 0.033 & 0.296 & 0.69 & 1.6 & 374.67/\textbf{374.67} & 369.29 & 3.84 & \textbf{10.0} & \textbf{0.206} & \textbf{369.4} & \textbf{-9.865}\\
        \hline
        \color{blue}1.968 & \color{blue}0.602 & \color{blue}1.249 & \color{blue}1.0 & \color{blue}0.0664 & \color{blue}0.374 & \color{blue}0.57 & \color{blue}2.0 & \color{blue}374.72/\textbf{374.67} & \color{blue}369.33 & \color{blue}3.82 & \color{blue}\textbf{8.9} & \color{blue}\textbf{0.213} & \color{blue}\textbf{369.4} & \color{blue}\textbf{-4.865}\\
        \hline
        1.97 & 0.58 & 1.25 & 1.0 & 0.032 & 0.283 & 0.6 & 2.1 & 374.73/\textbf{374.67} & 369.36 & 3.79 & \textbf{10.0} & \textbf{0.206} & \textbf{369.4} & \textbf{-1.865}\\
        \hline
        \color{blue}1.906 & \color{blue}0.524 & \color{blue}1.22 & \color{blue}1.0 & \color{blue}0.0655 & \color{blue}0.370 & \color{blue}0.8 & \color{blue}1.5 & \color{blue}374.67/\textbf{374.67} & \color{blue}369.34 & \color{blue}3.81 & \color{blue}\textbf{8.9} & \color{blue}\textbf{0.214} & \color{blue}\textbf{369.4} & \color{blue}\textbf{-0.865}\\
        \hline
        \color{blue}1.873 & \color{blue}0.51 & \color{blue}1.22 & \color{blue}1.0 & \color{blue}0.0668 & \color{blue}0.378 & \color{blue}0.7 & \color{blue}1.5 & \color{blue}374.63/\textbf{374.67} & \color{blue}369.32 & \color{blue}3.83 & \color{blue}\textbf{8.9} & \color{blue}\textbf{0.213} & \color{blue}\textbf{369.4} & \color{blue}\textbf{0.135}\\
        \hline
        \color{blue}1.876 & \color{blue}0.48 & \color{blue}1.19 & \color{blue}1.0 & \color{blue}0.0680 & \color{blue}0.370 & \color{blue}0.8 & \color{blue}1.1 & \color{blue}374.71/\textbf{374.67} & \color{blue}369.42 & \color{blue}3.75 & \color{blue}\textbf{8.9} & \color{blue}\textbf{0.214} & \color{blue}\textbf{369.4} & \color{blue}\textbf{1.135}\\
        \hline
        1.883 & 0.49 & 1.196 & 0.88 & 0.038 & 0.315 & 0.86 & 0.7 & 374.74/\textbf{374.67} & 369.36 & 3.79 & \textbf{10.0} & \textbf{0.208} & \textbf{369.4} & \textbf{5.135}\\
        \hline
        1.837 & 0.455 & 1.201 & 0.87 & 0.0432 & 0.298 & 0.61 & 1.0 & 374.60/\textbf{374.67} & 369.31 & 3.83 & \textbf{10.0} & \textbf{0.207} & \textbf{369.4} & \textbf{10.135}\\
        \hline
        1.82 & 0.39 & 1.14 & 0.74 & 0.036 & 0.282 & 0.5 & 2.5 & 374.58/\textbf{374.67} & 369.31 & 3.83 & \textbf{10.0} & \textbf{0.206} & \textbf{369.4} & \textbf{15.135}\\
        \hline

    \end{tabular}
    \caption{Fitted and fixed parameters (from top to bottom) corresponding to each data point (from left to right) in Fig.~5(a) .}
    \label{table : Fit Parameters fig5}
\end{table}

\begin{table}[htbp]
    \footnotesize
    \begin{tabular}{|c|c|c|c|c|c|c|c|c|c|c|c|c|c|c|}
        \hline
        \shortstack{$E_{+}$ \\ (V/m)} & \shortstack{$E_{-}$ \\ (V/m)} & \shortstack{$E_{0}$ \\ (V/m)} & $\phi/\pi$ & \shortstack{$E_{\textrm{DC}}$ \\ (V/m)} & OD & \shortstack{$\Gamma_{4}/2\pi$ \\ (MHz)} & \shortstack{$\Gamma_{3}/2\pi$ \\ (MHz)} & \shortstack{$\nu_p^\text{abs}$ \\ (MHz)} & \shortstack{$\nu_p^\text{act}$ \\ (MHz)} & \shortstack{$B_{\text{calc}}$ \\ (gauss)}  & \shortstack{$\bm{\Omega_{c}/2\pi}$ \\ \textbf{(MHz)}} & \shortstack{$\bm{\Omega_{p}/2\pi}$ \\ \textbf{(MHz)}} & \shortstack{$\bm{\nu_p^\textbf{park}}$ \\ \textbf{(MHz)}}  & \shortstack{$\bm{\Delta_m/2\pi}$ \\ \textbf{(MHz)}}\\ 
        \hline
        1.00 & 0.60 & 0.896 & 0.46 & 0.0473 & 0.415 & 0.78 & 0.2 & 374.01/\textbf{373.97} & 358.41 & 11.12 & \textbf{6.69} & \textbf{0.183} & \textbf{358.5} & \textbf{35}\\
        \hline
        0.649 & 0.845 & 1.000 & 0.44 & 0.0417 & 0.397 & 0.35 & 0.8 & 373.98/\textbf{373.97} & 365.39 & 6.13 & \textbf{6.8} & \textbf{0.180} & \textbf{365.2} & \textbf{6.5}\\
        \hline
        0.774 & 0.678 & 0.955 & 0.24 & 0.0539 & 0.403 & 0.50 & 0.61 & 373.95/\textbf{373.97} & 368.73 & 3.74 & \textbf{7.21} & \textbf{0.182} & \textbf{368.8} & \textbf{-23}\\
        \hline

    \end{tabular}
    \caption{Fitted and fixed parameters (from top to bottom) corresponding to each data point (from left to right) in Fig.~6.}
    \label{table : Fit Parameters fig6}
\end{table}

Table~\ref{table : Fit Parameters fig2} summarizes the data taken on 04/10/2026, which are used to examine multi-level nonlinearity. Table~\ref{table : Fit Parameters fig4} presents the results corresponding to the source-power-dependent measurements shown in Fig.~4 of the main text. The data are color-coded by acquisition date: blue (11/19/2025), black (11/20/2025), purple (11/21/2025), and brown (11/25/2025).

Table~\ref{table : Fit Parameters fig5} provides the results for the MW-frequency-dependent measurements shown in Fig.~5. These data were collected over two days: blue corresponds to 11/26/2025 and black to 11/27/2025. The data taken on 11/25/2025 in Table~\ref{table : Fit Parameters fig4} share the same fixed $\nu_p^\text{abs}$ as the data in this table.

Table~\ref{table : Fit Parameters fig6} shows the results for the MW-frequency-dependent measurements after introducing the stub tuner, corresponding to Fig.~6 in the main text. These measurements were performed on consecutive days (04/01–04/03/2026).

\subsection{Four-level}

The four-level microwave measurements shown in Fig.~5(a) of the main text are from a simplified fit process. The four-level fit model retains the dominant microwave coupling while neglecting the full Zeeman manifold structure, and therefore provides a comparison of the target MW component to the full multi-level analysis. It also ignores all the off-resonant couplings that are used in constructing the multi-level Hamiltonian, which will be discussed in Sec.~\ref{sec:off-resonant coupling}.

The free fitting parameters are the effective microwave Rabi frequency $\Omega_{\rm m}$, the residual dc electric field $E_{\rm DC}$, the optical depth ${\rm OD}$, the Rydberg state dephasing $\Gamma_3=\Gamma_4$, and the magnetic-field-shifted probe resonance frequency $\nu_p^\text{act}$:

\[
\{\Omega_{\rm m},E_{\rm DC},{\rm OD},\Gamma_3,\nu_p^\text{act}\}.
\]

Here, $\nu_p^\text{act}$ is still used to determine the magnetic field through Eq.~\ref{B_field}, and the control-laser detuning is constructed in the same way as Eq.~\ref{delta_c}.

The microwave detuning $\Delta_m$, control-laser Rabi frequency $\Omega_c$, and probe Rabi frequency $\Omega_p$ are treated as fixed parameters. The zero-field probe resonance $\nu_p^\text{abs}$ is also fixed to the average value obtained from the multi-level fits, since all four-level data were acquired on the same day or within a short time window of the multi-level measurements. As a result, only a single-round fitting procedure is performed for the four-level model.

In the four-level spectra, there is only a simple Autler–Townes Splitting $\Delta_\text{AT}$ structure, so also fit the two peaks as Gaussian functions for these near resonant 4-level spectra as a reference shown in the last column of Table~\ref{table : Fit Parameters 4level}.

\begin{table}[htbp]
    \footnotesize
    \begin{tabular}{|c|c|c|c|c|c|c|c|c|c|c|c|c||c|}
        \hline
        \shortstack{$E_{q}$ \\ (V/m)}  & \shortstack{$E_{\textrm{DC}}$ \\ (V/m)} & OD & \shortstack{$\Gamma_{3}/2\pi$ \\ (MHz)} & \shortstack{$\bm{\nu_p^\textbf{abs}}$ \\ \textbf{(MHz)}} & \shortstack{$\nu_p^\text{act}$ \\ (MHz)} & \shortstack{$B_{\text{calc}}$ \\ (gauss)}  & \shortstack{$\bm{\Omega_{c}/2\pi}$ \\ \textbf{(MHz)}} & \shortstack{$\bm{\Omega_{p}/2\pi}$ \\ \textbf{(MHz)}} & \shortstack{$\bm{\nu_p^\textbf{park}}$ \\ \textbf{(MHz)}}  & \shortstack{$\bm{\Delta_m/2\pi}$ \\ \textbf{(MHz)}} &
        \shortstack{$\delta_{m}/2\pi$ \\ (MHz)}  &
        \shortstack{$\Omega_{m}/2\pi$ \\ (MHz)}  &
        \shortstack{$\Delta_{\text{AT}}/2\pi$ \\ (MHz)} \\
        \hline
         0.128 & 0.074 & 0.392 & 1.06 & \textbf{374.67} & 348.20 & 18.91 & \textbf{4.0} & \textbf{0.229} &  \textbf{348.5} & \textbf{-28.265} & 0.14 & 3.6(1) & 3.77(9) \\
        \hline
         0.237 & 0.056 & 0.403 & 0.95 & \textbf{374.67} & 348.16 & 18.94 & \textbf{4.0} & \textbf{0.230} &  \textbf{348.5} & \textbf{8.135} & 1.57 & 5.4(1) & 5.86(9)\\
        \hline
         \color{blue}0.066 & \color{blue}0.054 & \color{blue}0.247 & \color{blue}0.60 & \color{blue}\textbf{374.67} & \color{blue}356.50 & \color{blue}12.98 & \color{blue}\textbf{3.1} & \color{blue}\textbf{0.204} &  \color{blue}\textbf{356.6} & \color{blue}\textbf{28.735} & \color{blue}0.45 & \color{blue}1.07(5) & \color{blue}1.26(7)\\
        \hline
         \color{blue}0.111 & \color{blue}0.050 & \color{blue}0.221 & \color{blue}0.34 & \color{blue}\textbf{374.67} & \color{blue}352.74 & \color{blue}15.67 & \color{blue}\textbf{3.1} & \color{blue}\textbf{0.203} &  \color{blue}\textbf{352.8} & \color{blue}\textbf{35.735} & \color{blue}1.00 & \color{blue}1.81(6) & \color{blue}2.13(4)\\
        \hline
         0.279 & 0.050 & 0.398 & 1.2 & \textbf{374.67} & 348.35 & 18.81 & \textbf{4.0} & \textbf{0.230} &  \textbf{348.5} & \textbf{41.735} & -0.31 & 4.5(1) & 4.53(9)\\
        \hline
         \color{brown}0.127 & \color{brown}0.026 & \color{brown}0.379 & \color{brown}0.34 & \color{brown}\textbf{373.97} & \color{brown}352.4 & \color{brown}15.37 & \color{brown}\textbf{3.0} & \color{brown}\textbf{0.184} &  \color{brown}\textbf{352.2} & \color{brown}\textbf{35} & \color{brown}-0.30 & \color{brown}2.07(7) & \color{brown}2.26(5) \\
        \hline
         \color{brown}0.135 & \color{brown}0.039 & \color{brown}0.399 & \color{brown}0.32 & \color{brown}\textbf{373.97} & \color{brown}351.9 & \color{brown}15.76 & \color{brown}\textbf{2.9} & \color{brown}\textbf{0.181} &  \color{brown}\textbf{352.4} & \color{brown}\textbf{6.5} & \color{brown}0.29 & \color{brown}3.10(6) & \color{brown}3.12(6)\\
        \hline
         \color{brown}0.097 & \color{brown}0.061 & \color{brown}0.405 & \color{brown}0.48 & \color{brown}\textbf{373.97} & \color{brown}351.95 & \color{brown}15.73 & \color{brown}\textbf{3.0} & \color{brown}\textbf{0.183} &  \color{brown}\textbf{352.4} & \color{brown}\textbf{-23} & \color{brown}0.84 & \color{brown}2.73(7) & \color{brown}2.90(7)\\
        \hline

    \end{tabular}
    \caption{Fitted and fixed parameters (from top to bottom) corresponding to each 4-level data point (from left to right) in Fig.~5(a) and Fig.~6.}
    \label{table : Fit Parameters 4level}
\end{table}

Table~\ref{table : Fit Parameters 4level} presents the results corresponding to all the 4-level measurements shown in Fig.~5 and Fig.~6 of the main text. The data are color-coded by acquisition date: black (11/24/2025), blue (11/28/2025), and brown (04/01–04/03/2026).

 \section{off-resonant coupling to additional Rydberg states}
\label{sec:off-resonant coupling}
    \begin{figure}[ht]
\centering
\includegraphics[width=\textwidth]{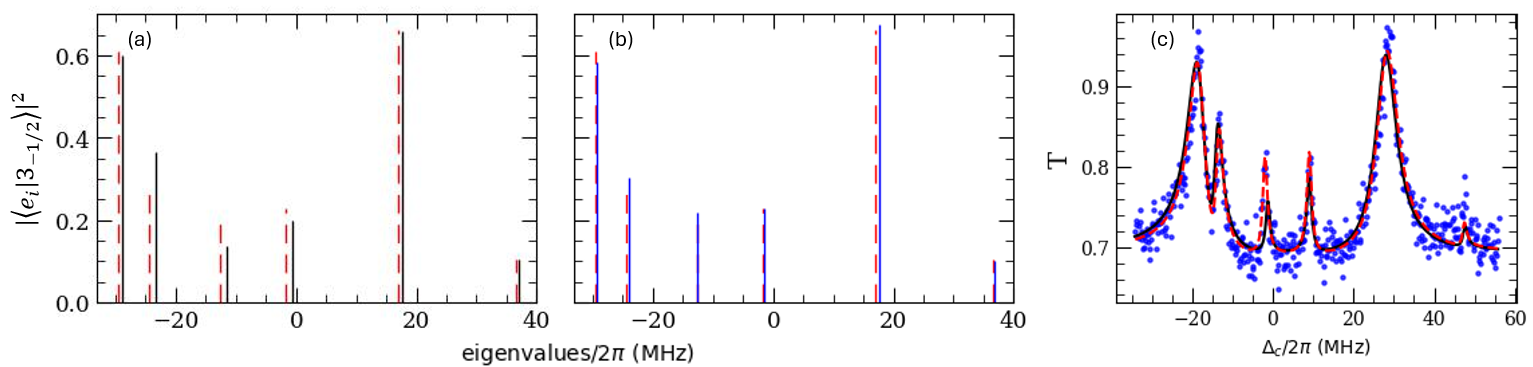}
\caption{The effect of off-resonant couplings to additional Rydberg states. (a, b) The eigenspectrum of the 6-by-6 Hamiltonian involving only the Rydberg levels. The bar height represents the overlap of the corresponding eigenstate and $\ket{3_{-1/2}}$. In (a), the dashed (solid) lines are obtained by solving the Hamiltonian including the perturbative contribution only from $62S$(without contributions from any off-resonant levels). In (b), the dashed (solid) lines include contributions from $62S$ (both $62S$ and $61P_{1/2}$). (c) The MW EIT spectroscopy along with the fits excluding off-resonant couplings (solid black line) and including couplings to both $62S$ and $61P_{1/2}$ perturbatively (dashed red line). 
}
\label{fig:off resonant coupling}
\end{figure}

We consider off-resonant couplings to other Rydberg states in addition to the target states $\ket{61S_{1/2}}$ and $\ket{61P_{3/2}}$. In the absence of external fields, the transition $\ket{61P_{3/2}}\leftrightarrow \ket{62S_{1/2}}$ is 218~MHz above the target transition $\ket{61S_{1/2}}\leftrightarrow\ket{61P_{3/2}}$. This is the transition closest to the target one. The next closest transition is $\ket{61S_{1/2}}\leftrightarrow\ket{61P_{1/2}}$, which is 437~MHz below the target transition.

We diagonalize the Hamiltonian including MW couplings in the Rydberg manifold and plot the eigenenergies in Fig.~\ref{fig:off resonant coupling}. Example parameters are from Table~\ref{table : Fit Parameters fig5} $\Delta_m/(2\pi)=-0.865$~MHz, corresponding to the multi-level spectrum in Fig.~5(a) of the main text (from left to right). 
We first verify that a perturbative effective Hamiltonian treatment of the off-resonant states gives nearly identical results to diagonalizing the full Hamiltonian. In this perturbative treatment, the off-resonant couplings introduce AC Stark shifts of the target Rydberg states and Raman couplings between different Zeeman sublevels within the target Rydberg states that otherwise would not be coupled. The motivation for this treatment is the time cost for the fit: including many more states would significantly slow down the fitting procedure.

We compare eigenenergies and projection of the corresponding eigenstates to $\ket{3_{-1/2}}$ obtained by including only the intended states versus adding $\ket{62S_{1/2}}$ (Fig.~\ref{fig:off resonant coupling}(a)). Including $\ket{62S_{1/2}}$ leads to eigenenergy shifts of up to roughly 2~MHz, along with appreciable changes in state overlaps. Next we compare the results obtained by including only $\ket{62S_{1/2}}$ versus including both $\ket{62S_{1/2}}$ and $\ket{61P_{1/2}}$ (Fig.~\ref{fig:off resonant coupling}(b)). The impact of coupling to $\ket{61P_{1/2}}$ is very small. Therefore, we do not consider other states with much larger detunings. Furthermore, we study possible two-photon off-resonant couplings from the intended states and confirm that none is close to resonance.

In Fig.~\ref{fig:off resonant coupling}(c), we show the two fit results for the 6th multi-level dataset in Fig.~5(a) of the main text (from left to right), obtained by including only the target Rydberg states and by perturbatively including off-resonant couplings to $\ket{62S_{1/2}}$ and $\ket{61P_{1/2}}$. The differences in the extracted MW electric-field amplitudes are $2$\% for $E_{+}$, $5$\% for $E_0$, and $15$\% for $E_{-}$, along with a $0.017\pi$ difference in MW phase $\phi$. 

$E_0$ and $E_{-}$ are outside the fit uncertainties and the fit obtained from the model not involving off-resonant couplings is a visibly poor fit (solid black line in Fig.~\ref{fig:off resonant coupling}(c)), where the third peak from the left is misaligned with the data. Therefore, we include these additional states in all multi-level fits, although they matter appreciably only for some datasets.

\section{Phase sensitivity}
\label{sec:D to P transition phase sensitivity}

\begin{figure}[ht]
\centering
\includegraphics[width=\textwidth]{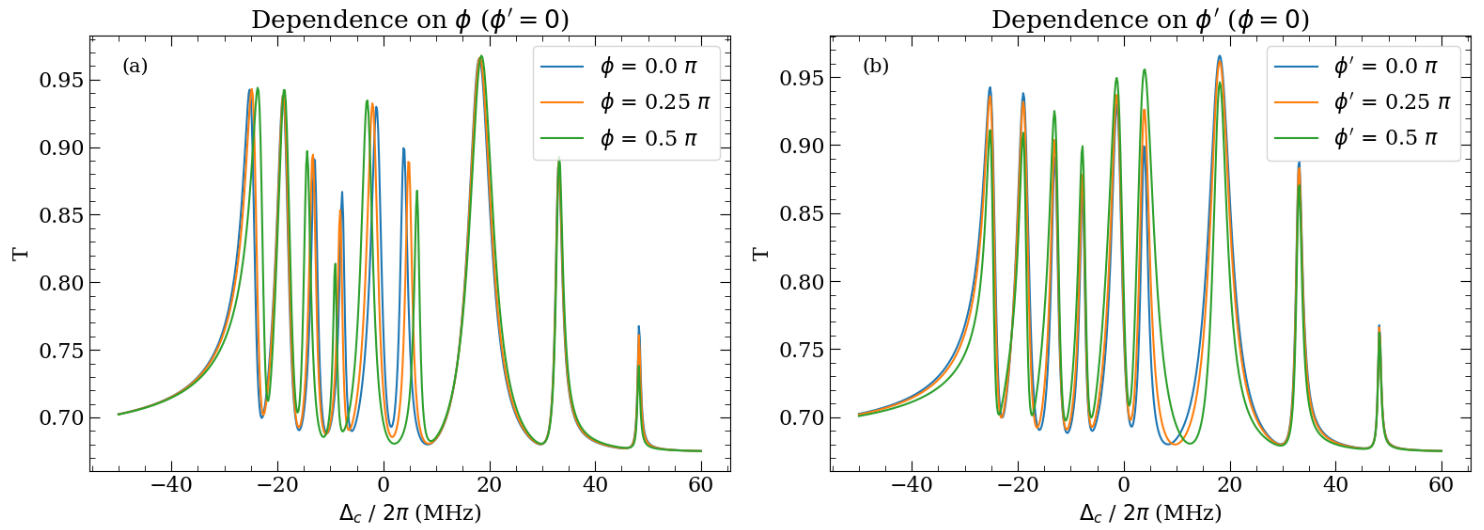}
\caption{MW phase sensitivity of spectra of $\ket{D_{5/2}}\leftrightarrow\ket{P_{3/2}}$ transition. Conditions are set the same as the center dataset in Fig.~6(a-c) of the main text except the MW frequency. (a) $\phi$ sensitivity. (b) $\phi'$ sensitivity.}
\label{fig:phase sensitivity}
\end{figure}

Using the $\ket{D_{5/2}}\leftrightarrow\ket{P_{3/2}}$ transition and a linearly polarized control field, the spectrum is determined by two interferometric loops as shown in Fig.~3 of the main text. Therefore, the spectrum is sensitive to two phase combinations $\phi=2\phi_0-\phi_+-\phi_-$ and $\phi'=\phi_+-\phi_--\phi_c$, where $\phi_c$ is the relative phase between the $\sigma^+$ and $\sigma^-$ control polarization components. Similar to the $\ket{S_{1/2}}\leftrightarrow\ket{P_{3/2}}$ transition, $\phi$ and $\phi'$ have a four-fold ambiguity within a full period of $2\pi$. In other words, four phase pairs $(\phi,\phi')$, which generally correspond to distinct 3D vector MW electric fields, may share the same spectrum.

We numerically simulate the model using parameters identical to the middle multi-level dataset in Fig.~6(a-c) of the main text, except that the MW frequency is now resonant with the $\ket{60D_{5/2}}\leftrightarrow\ket{61P_{3/2}}$ transition at 9.7077~GHz. In the simulation (Fig.~\ref{fig:phase sensitivity}), the $\sigma^+$ and $\sigma^-$ control Rabi frequencies are set equal, rather than the electric-field amplitudes, which would correspond to linear polarization. This choice balances the effective coupling strengths and improves the phase sensitivity.

In practice, this requires an elliptically polarized control field, with the ratio of the $\sigma^-$ over $\sigma^+$ control electric field amplitude 
\begin{equation}
r=\frac{|E^c_-|}{|E^c_+|}=0.3
\end{equation}
Then the control field ellipticity magnitude is
\begin{equation}
\epsilon_c=\frac{\left||E^c_+|-|E^c_-|\right|}{|E^c_+|+|E^c_-|}
=\frac{|1-r|}{1+r}
\end{equation}

The appearance of the control relative phase $\phi_c$ is rooted in the system's rotational symmetry about the quantization axis. If the probe field is not purely $\sigma^-$ polarized, the relative phase $\phi_p$ between its polarization components would similarly appear in a closed-loop phase combination and serve as a frame reference equivalent to $\phi_c$. However, mixing probe polarizations drives transitions away from the strongest $\ket{1}\leftrightarrow\ket{2}$ cycling transition, reducing absorption and therefore spectral contrast. 
This rotational symmetry could instead be removed by repeating the measurement with two nonparallel quantization axes.
In this two-axis implementation, the reconstructed MW vector field is referenced to the calibrated geometry of the bias magnetic fields.
We leave a systematic study of schemes using additional probe polarization components and the two-axis approach to future work.

Both approaches avoid calibration of external reference MW fields and thus preserve the self-calibrated character of the measurement. If the reference is provided by an additional optical polarization component, the required calibration is a one-time characterization of the beam polarization using standard polarization optics.
In the two-axis implementation, the relative directions of the applied bias fields can be calibrated in situ using the atoms themselves. Specifically, measuring the magnetic field magnitude via Zeeman spectroscopy for many coil-current combinations and fitting the resulting quadratic form determines the cross terms, which give the dot products between the actual field directions at the atoms.

\section{Fit uncertainty}
\begin{figure*}[ht]
\centering
\includegraphics[width=\textwidth]{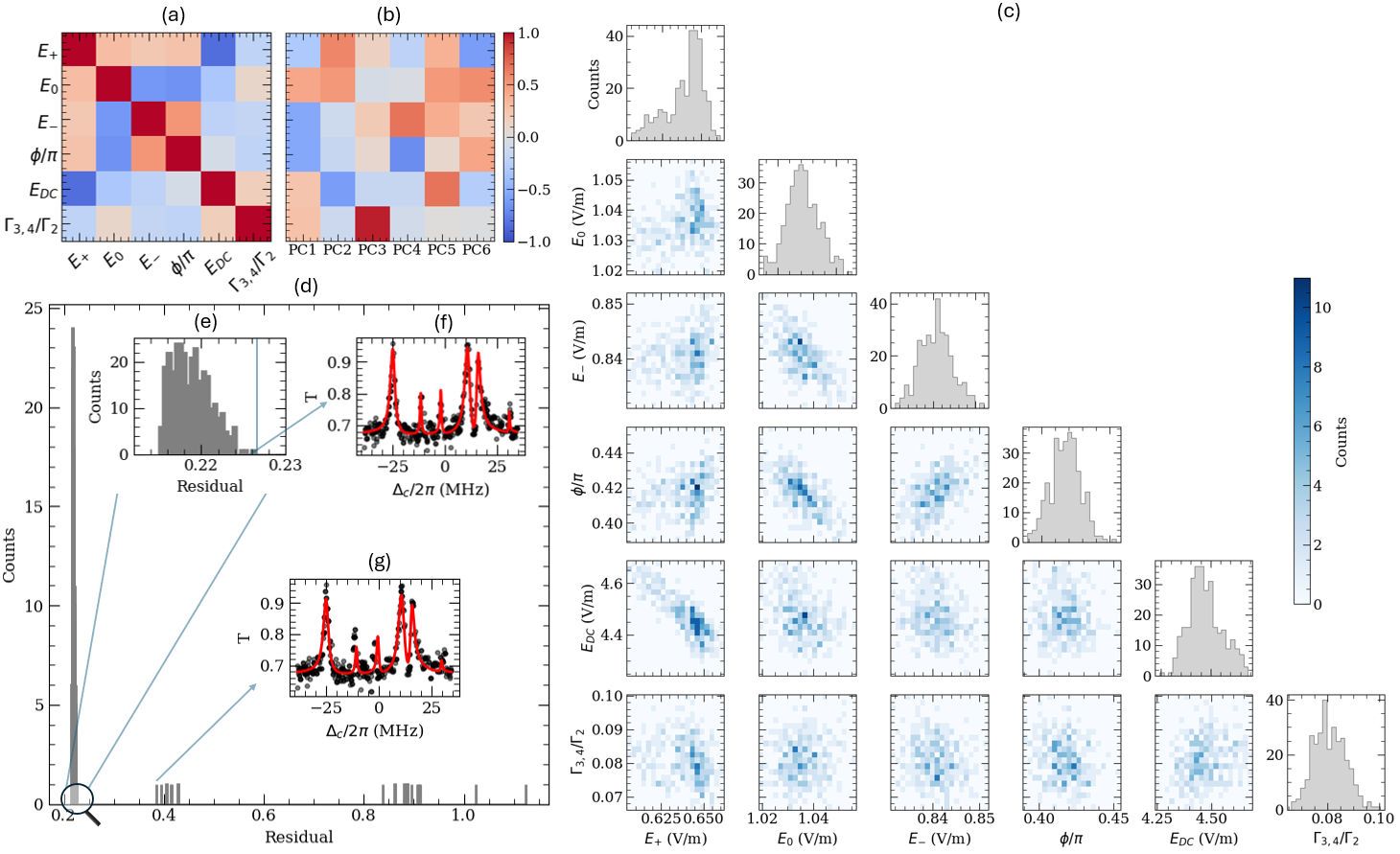}
\caption{Non-parametric bootstrap result. The correlation (a) and the loading (b) matrices for the 301 non-parametric bootstrap replicas corresponding to the middle multi-level dataset in Fig.~6(a-c) of the main text. PC\textit{i} refers to the \textit{i}th Principal Component. (c) Corner plot showing the 1D histograms and pair-wise 2D scatter plots of the fit parameters. (d) Histogram of the residuals. (e) Zoomed-in of (d) around Residual = 0.22, with the blue line indicating the cutoff threshold above which bootstrap replicas are discarded as visually poor fits. (f) Data (dots) and fit (solid line) corresponding to a residual of 0.22631, the largest value shown in (e). (g) Data and fit corresponding to a residual of 0.384, the smallest residual above the cutoff; in this case, two of the six peaks are visibly misaligned with the data. Here, OD is fixed at the value obtained from curvefit for faster computation.}
\label{fig:fit uncertainty}
\end{figure*}
The covariance matrix returned by the nonlinear least-squares routine is only strictly valid when the model is approximately linear in all fit parameters near the optimum. Given the large number of parameters in our model, these conditions are not always satisfied. To obtain more reliable uncertainty estimates, we therefore perform both nonparametric case-resampling bootstrap and parametric bootstrap analyses~\cite{davison1997bootstrap,hesterberg2011bootstrap}. In this section, we use the set of parameters corresponding to $\Delta_m/(2\pi)=6.5$~MHz from Table~\ref{table : Fit Parameters fig6}.

The default bootstrap method is the nonparametric case-resampling bootstrap: each measured data point ($\Delta_c^i,T^i$) is treated as a pair, and we repeatedly draw a same-size sample with replacement from the original spectrum and refit the full model. The resulting distribution of fitted parameters yields empirical uncertainties and correlations. 

For each bootstrap sample, the refit is performed in two steps. First, we apply a  global search using \texttt{differential\_\allowbreak evolution}. There, the maximum iteration is set to 20 and fit candidate population multiplier is 25 for the balance of speed and initial guess free. Mutation interval is set to (0.5, 1) and recombination to 0.9 to avoid local minima while ensuring it converges relatively fast. The relative tolerance is 1$\%$. Then, the fit result is polished with the \texttt{curve\_fit} function. Both the fit functions are from the \texttt{scipy.optimize} package. Parallelization is enabled in the global search stage to accelerate the process and each bootstrap replica takes roughly 30 minutes to fit. 

\begin{figure*}[ht]
\centering
\includegraphics[width=\textwidth]{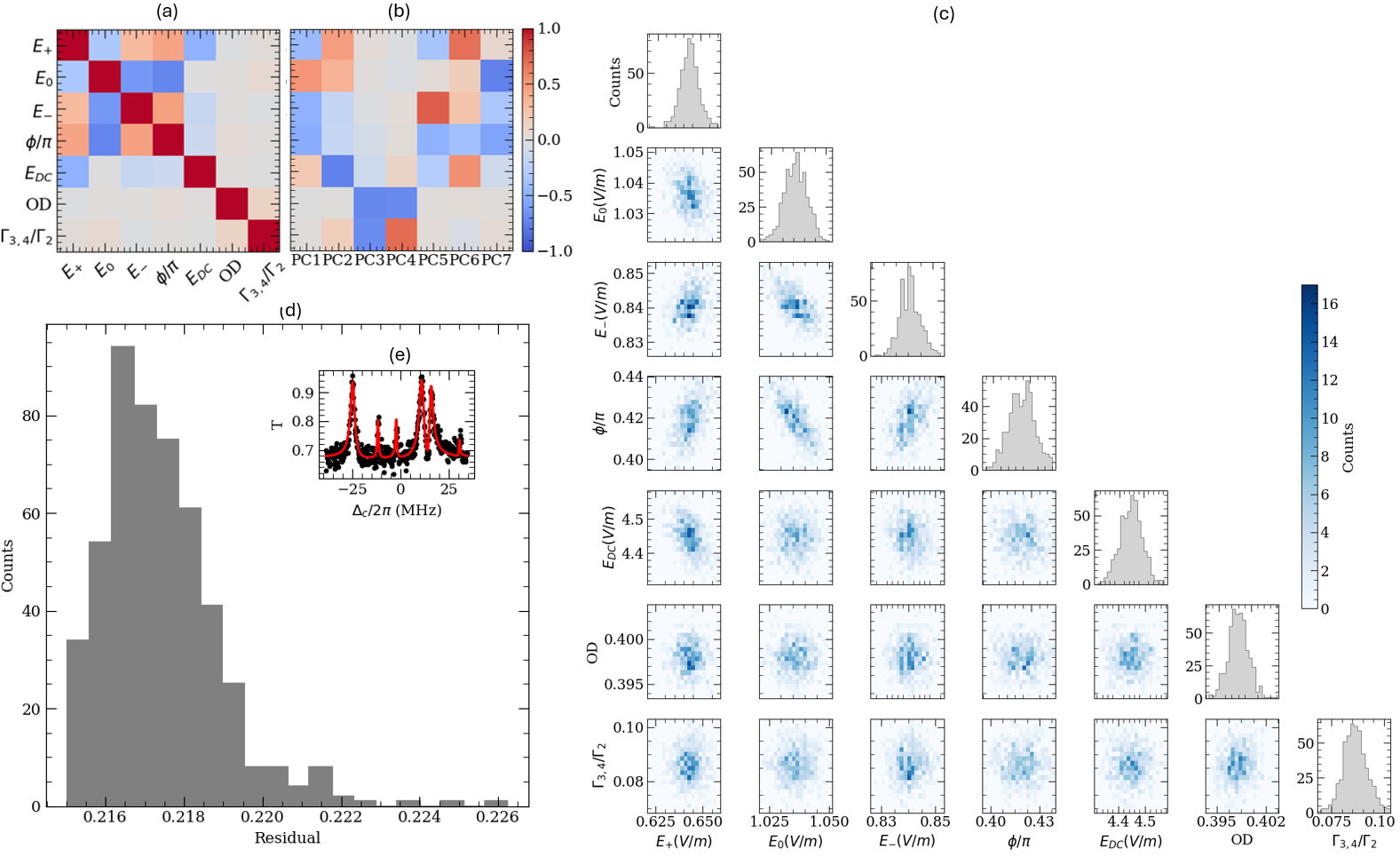}
\caption{Parametric bootstrap result. The correlation (a) and the loading (b) matrices for the 500 parametric bootstrap replicas corresponding to the middle multi-level dataset in Fig.~6(a-c) of the main text. PC\textit{i} refers to the \textit{i}th Principal Component. (c) Corner plot showing the 1D histograms and pair-wise 2D scatter plots of the fit parameters. (d) Histogram of the residuals. (e) Data (dots) and fit (solid line) corresponding to a residual of 0.2262, the largest value among all the 500 replicas. As the fit looks visually good, we do not reject any replica to construct (a),(b) and (c).}
\label{fig:parametric bootstrap}
\end{figure*}

These parameters are tuned to maximize the fraction of bootstrap replicas that converge to the low-residual solution branch shown in Fig.~\ref{fig:fit uncertainty}(d).
We create 319 bootstrap replicas and compute the residual of each fit, defined as the sum of squared differences between the original data and the fitted model (Fig.~\ref{fig:fit uncertainty}(d,e)). We disregard results with residuals outside the low-residual cluster, based on the visual alignment of the peaks in the fit with the data (Fig.~\ref{fig:fit uncertainty}(f,g)). For this dataset, the cutoff is 0.2264, leaving us with 301 replicas. The correlation and loading matrices, as well as the corner plot, are created from these 301 samples. (Fig.~\ref{fig:fit uncertainty}(a,b,c)).
Additionally, there is an initial set of parameters for the fitting process that has the same order of magnitude as the final results, but is not specifically tuned to the dataset. We observe an asymmetric distribution of the fitted $E_{+}$ and a large negative correlation between $E_{+}$ and $E_{\text{DC}}$. However, we do not believe that these features represent the true information in either the data or the model. Refitting with $E_{+}$ constrained to values on either side of the median returns nearly identical residuals. Additionally, the parametric bootstrap returns a symmetric $E_{+}$ distribution and a moderate correlation between $E_{+}$ and $E_{\text{DC}}$ (Fig.~\ref{fig:parametric bootstrap}). 

In the parametric bootstrap method, we assume our data follows a distribution described by the fit model with Gaussian noise on each data point. The Gaussian noise is generated in transmission and added to the fit result. The noise level is chosen such that it characterizes actual data the best.
In the dataset presented in Figs.~\ref{fig:fit uncertainty} and \ref{fig:parametric bootstrap}, all three methods--covariance from a single least square fit, 68\% confidence interval from nonparametric and parametric bootstrap all give results statistically in agreement (Table~\ref{table:Bootstrap}).   
\begin{table}[htbp]
    \centering
    \footnotesize
    \begin{tabular}
    {|c|c|c|c|c|}
    \hline
    Method & $E_{+}$ (V/m) & $E_{0}$ (V/m)& $E_{-}$ (V/m)& $\phi/\pi$ \\
    \hline
    Curvefit & 0.643(6) & 1.037(6) & 0.841(4) & 0.42(1) \\
    \hline
    Non-parametric Bootstrap & $0.643^{+0.007}_{-0.018}$& $1.035^{+0.008}_{-0.006}$ & $0.841^{+0.003}_{-0.004}$ & $0.418^{+0.009}_{-0.012}$ \\
    \hline
    Parametric Bootstrap & $0.643^{+0.005}_{-0.005}$ & $1.035^{+0.005}_{-0.005}$ & $0.841^{+0.004}_{-0.004}$ & $0.420^{+0.008}_{-0.009}$ \\
    \hline
    \end{tabular}
    
    \caption{The MW parameters and their errorbars from the three methods - the covariance from the least square fit (curvefit) of the raw data, 68\% confidence interval from non-parametric  and parametric bootstrap - for the middle multi-level dataset in Fig.~6(a-c) of the main text. }
    \label{table:Bootstrap}
\end{table}

To reduce the computational cost, the non-parametric bootstrap, which is the most time-consuming process out of  the three methods, treats OD as a fixed parameter. This is justified as OD is weakly correlated to the other parameters as evident from the parametric bootstrap (Fig.~\ref{fig:parametric bootstrap}(a)).

\section{MW delivery}

For the data in Figs.~4 and 5 in the main text, the MW horn (FMWAN159-10SF, Fairview Microwave) is mounted approximately 20~cm from the atoms and is oriented toward the atomic cloud at an angle of 90$^\circ$ relative to the quantization axis. The MW source (Anritsu 68369A/NV) is connected to a directional coupler (Anritsu D29422) and then to the horn.

\begin{figure}[ht]
\centering
\includegraphics[width=1\textwidth]{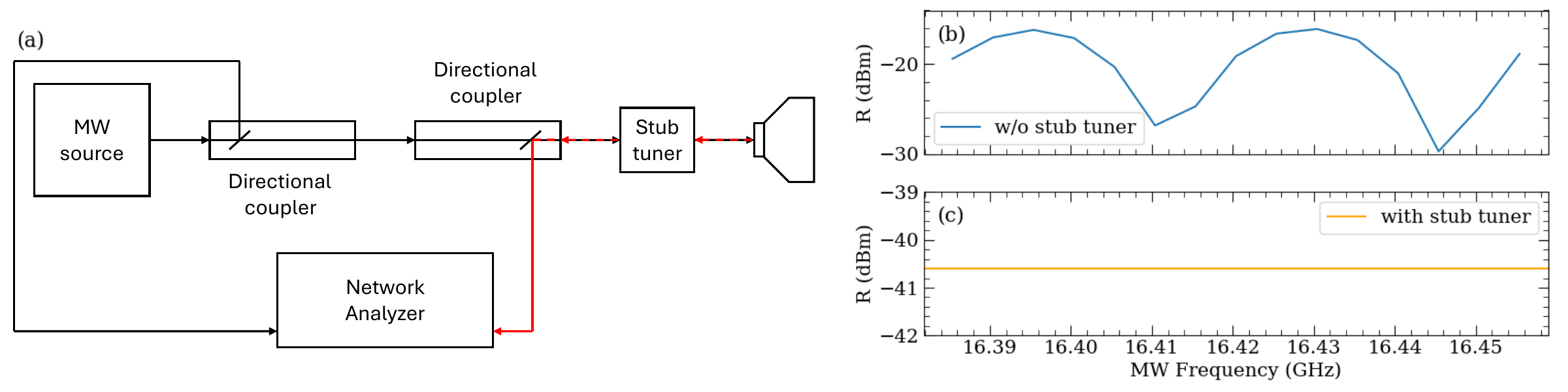}
\caption{ Microwave impedance match. (a)  Setup for MW delivery. The first directional coupler (16~dB) samples the source and the second one (10~dB) samples the reflection. (b) Reflected microwave power vs frequency without stub tuner. (c) Reflected microwave power is maintained a constant for each frequency.}
\label{fig:MW circuit and impedance matching}
\end{figure}

For the data in Figs.~2 and 6 of the main text, the MW horn is placed at a different position and is impedance matched as shown in Fig.~\ref{fig:MW circuit and impedance matching}(a). The MW horn is mounted approximately 40~cm from the atoms and is oriented toward the atomic cloud at an angle of 45$^\circ$ relative to the quantization axis. The horn is mounted on a rotation stage, which allows the horn angle to be varied and read out with an uncertainty of about $1^\circ$. The MW source is connected to two directional couplers (Narda 4203-16 and Narda 4203-10) in sequence, followed by a stub tuner (Maury Microwave 9556A) and subsequently to the MW horn via an SMA cable (CBL-10FT-SMSM+, Minicircuits). The first directional coupler, which is closer to the source, monitors the forward power sent to the cable and horn assembly. The second directional coupler is used to monitor the reflection from the cable and horn assembly. Both signals are measured using a network analyzer (Anritsu 56100A). The forward-power pickoff measured at the first directional coupler is constant at $-6.43$~dBm. In the absence of the stub tuner, the reflected power $R$ measured at the second directional coupler varies by as much as $14$~dB over $10$~MHz, as shown in Fig.~\ref{fig:MW circuit and impedance matching}(b). We therefore adjust the stub tuner at each frequency to minimize $R$, which keeps the reflected-power pickoff low at about $-40.6$~dBm, across the frequency range used for our MW sensing experiments, as shown in Fig.~\ref{fig:MW circuit and impedance matching}(c).

\bibliography{references}

@article{chilcott2026quantum,
  title={Quantum-enabled complete RF-polarimetry with an optically-wired atomic sensor},
  author={Chilcott, Matthew and Stokholm, Laurits N and Cloutman, Matthew and Otto, J Susanne and Deb, Amita B and Kj{\ae}rgaard, Niels},
  journal={arXiv preprint arXiv:2605.14529},
  year={2026}
}

@article{PhysRevA.109.L021702,
  title = {Zeeman-resolved Autler-Townes splitting in Rydberg atoms with tunable resonances and a single transition dipole moment},
  author = {Schlossberger, Noah and Rotunno, Andrew P. and Artusio-Glimpse, Alexandra B. and Prajapati, Nikunjkumar and Berweger, Samuel and Shylla, Dangka and Simons, Matthew T. and Holloway, Christopher L.},
  journal = {Phys. Rev. A},
  volume = {109},
  issue = {2},
  pages = {L021702},
  numpages = {6},
  year = {2024},
  month = {Feb},
  publisher = {American Physical Society},
  doi = {10.1103/PhysRevA.109.L021702},
  url = {https://link.aps.org/doi/10.1103/PhysRevA.109.L021702}
}

@article{xue2004phase,
  title={The phase dependent double electromagnetically induced transparency in a four-level system with closed interaction contour},
  author={Xue, Yan and Wang, Gang and Wu, Jin-Hui and Xu, Wei-Hua and Wang, Hai-Hua and Gao, Jin-Yue and Babin, SA},
  journal={Physics Letters A},
  volume={324},
  number={5-6},
  pages={388--395},
  year={2004},
  publisher={Elsevier}
}

@incollection{arimondo2018laser,
  title={Laser phase spectroscopy in closed-loop multilevel schemes},
  author={Arimondo, Ennio},
  booktitle={Exploring the World with the Laser: Dedicated to Theodor H{\"a}nsch on his 75th birthday},
  pages={665--677},
  year={2018},
  publisher={Springer}
}

@article{hesterberg2011bootstrap,
  title={Bootstrap},
  author={Hesterberg, Tim},
  journal={Wiley Interdisciplinary Reviews: Computational Statistics},
  volume={3},
  number={6},
  pages={497--526},
  year={2011},
  publisher={Wiley Online Library}
}

@article{glick2026low,
  title={Low frequency electric field sensing with a Rydberg beam},
  author={Glick, Jeremy and Dickson, John R and Wood, Josie and Kunz, Paul},
  journal={arXiv preprint arXiv:2604.01513},
  year={2026}
}

@article{Hammerland2026,
  url = {https://arxiv.org/abs/2603.06935},
  author = {Hammerland,  Daniel and Talashila,  Rajavardhan and Manchaiah,  Dixith and Prajapati,  Nikunjkumar and Schlossberger,  Noah and McKee,  Erik and Highman,  Michael A. and Simons,  Matthew T. and Berweger,  Samuel and Artusio-Glimpse,  Alexandra B. and Holloway,  Christopher L.},
  title = {MHz to sub-kHz field detection with an all-dielectric potassium Rydberg-atom sensor},
  journal={arXiv preprint arXiv:2603.06935},
  year = {2026},
  copyright = {arXiv.org perpetual,  non-exclusive license}
}

@article{wang2026sensing,
  title={Sensing Low-Frequency Field with Rydberg Atoms via Quantum Weak Measurement},
  author={Wang, Ding and Jin, Shenchao and Fan, Xiayang and Li, Hongjing and Liu, Jiatian and Huang, Jingzheng and Zeng, Guihua and Sun, Yuan},
  journal={arXiv preprint arXiv:2603.09518},
  year={2026}
}

@article{chandra2026electrometry,
  title={Electrometry of extremely-low frequencies from kHz to sub-Hz with a Rydberg-atom sensor},
  author={Chandra, Aveek and Paensin, Narongrit and Dumke, Rainer},
  journal={arXiv preprint arXiv:2603.13827},
  year={2026}
}

@article{kaina2014shaping,
  title={Shaping complex microwave fields in reverberating media with binary tunable metasurfaces},
  author={Kaina, Nad{\`e}ge and Dupr{\'e}, Matthieu and Lerosey, Geoffroy and Fink, Mathias},
  journal={Scientific reports},
  volume={4},
  number={1},
  pages={6693},
  year={2014},
  publisher={Nature Publishing Group UK London}
}

@book{davison1997bootstrap,
  title={Bootstrap methods and their application},
  author={Davison, Anthony Christopher and Hinkley, David Victor},
  number={1},
  year={1997},
  publisher={Cambridge university press}
}

@article{shu2024eliminating,
  title={Eliminating Incoherent Noise: A Coherent Quantum Approach in Multi-Sensor Dark Matter Detection},
  author={Shu, Jing and Xu, Bin and Xu, Yuan},
  journal={arXiv preprint arXiv:2410.22413},
  year={2024}
}

@article{chen2024quantum,
  title={Quantum enhancement in dark matter detection with quantum computation},
  author={Chen, Shion and Fukuda, Hajime and Inada, Toshiaki and Moroi, Takeo and Nitta, Tatsumi and Sichanugrist, Thanaporn},
  journal={Physical Review Letters},
  volume={133},
  number={2},
  pages={021801},
  year={2024},
  publisher={APS}
}

@article{chen2026background,
  title={Background suppression in quantum sensing of dark matter via collective entangled-state projection},
  author={Chen, Shion and Fukuda, Hajime and Iiyama, Yutaro and Mino, Yuya and Moroi, Takeo and Nakahara, Mikio and Nitta, Tatsumi and Sichanugrist, Thanaporn},
  journal={Physical Review D},
  volume={113},
  number={5},
  pages={055014},
  year={2026},
  publisher={APS}
}

@article{wang2026quantum,
  title={Quantum enhanced metrology based on flipping trajectory of cold Rydberg gases},
  author={Wang, Ya-Jun and Zhang, Jun and Zhang, Zheng-Yuan and Shao, Shi-Yao and Li, Qing and Chen, Han-Chao and Ma, Yu and Han, Tian-Yu and Wang, Qi-Feng and Nan, Jia-Dou and others},
  journal={Nature Communications},
  year={2026},
  publisher={Nature Publishing Group UK London}
}

@article{palm2026enhanced,
  title={Enhanced Rydberg Blockade through RF-tuned F$\backslash$" orster Resonance},
  author={Palm, Lukas and Li, Bowen and Feng, Yiming Cady and J{\"u}rgensen, Marius and Simon, Jon},
  journal={arXiv preprint arXiv:2603.07958},
  year={2026}
}

@article{wang2026nonlinear,
  title={Nonlinear optical spectra from Rydberg-mediated photon-photon interactions},
  author={Wang, Xinghan and Wang, Yupeng and Panja, Aishik and Liang, Qi-Yu},
  journal={arXiv preprint arXiv:2602.11563},
  year={2026}
}

@article{zhang2025microwave,
  title={Microwave electrometry with quantum-limited resolutions in a Rydberg atom array},
  author={Zhang, Yao-Wen and Xiang, De-Sheng and Liao, Ren and Liu, Hao-Xiang and Xu, Biao and Zhou, Peng and Zhou, Yijia and Zhang, Kuan and Li, Lin},
  journal={arXiv preprint arXiv:2512.05413},
  year={2025}
}

@article{kitching2025atom,
  title={Atom-based quantum sensing of electromagnetic fields},
  author={Kitching, John and Shaffer, James P and Budker, Dmitry},
  journal={Optica},
  volume={12},
  number={12},
  pages={2008--2022},
  year={2025},
  publisher={Optica Publishing Group}
}

@article{vendeiro2022machine,
  title={Machine-learning-accelerated Bose-Einstein condensation},
  author={Vendeiro, Zachary and Ramette, Joshua and Rudelis, Alyssa and Chong, Michelle and Sinclair, Josiah and Stewart, Luke and Urvoy, Alban and Vuleti{\'c}, Vladan},
  journal={Physical Review Research},
  volume={4},
  number={4},
  pages={043216},
  year={2022},
  publisher={APS}
}

@article{adams2019rydberg,
  title={Rydberg atom quantum technologies},
  author={Adams, Charles S and Pritchard, Jonathan D and Shaffer, James P},
  journal={Journal of Physics B: Atomic, Molecular and Optical Physics},
  volume={53},
  number={1},
  pages={012002},
  year={2019},
  publisher={IOP Publishing}
}

@article{wang2026robust,
  title={Robust Rydberg facilitation via rapid adiabatic passage},
  author={Wang, Xinghan and Wang, Yupeng and Liang, Qi-Yu},
  journal={Physical Review Research},
  volume={8},
  number={1},
  pages={013154},
  year={2026},
  publisher={APS}
}

@article{zhou2023improving,
  title={Improving the spectral resolution and measurement range of quantum microwave electrometry by cold Rydberg atoms},
  author={Zhou, Fei and Jia, Fengdong and Liu, Xiubin and Yu, Yonghong and Mei, Jiong and Zhang, Jian and Xie, Feng and Zhong, Zhiping},
  journal={Journal of Physics B: Atomic, Molecular and Optical Physics},
  volume={56},
  number={2},
  pages={025501},
  year={2023},
  publisher={IOP Publishing}
}

@article{liao2020microwave,
  title={Microwave electrometry via electromagnetically induced absorption in cold Rydberg atoms},
  author={Liao, Kai-Yu and Tu, Hai-Tao and Yang, Shu-Zhe and Chen, Chang-Jun and Liu, Xiao-Hong and Liang, Jie and Zhang, Xin-Ding and Yan, Hui and Zhu, Shi-Liang},
  journal={Physical Review A},
  volume={101},
  number={5},
  pages={053432},
  year={2020},
  publisher={APS}
}

@article{jamieson2025continuous,
  title={Continuous time ultra-high frequency (UHF) sensing using ultra-cold Rydberg atoms},
  author={Jamieson, Matthew J and Weatherill, Kevin J and Adams, C Stuart and Hanley, Ryan K and Alves, Natalia and Keaveney, James},
  journal={arXiv preprint arXiv:2504.00212},
  year={2025}
}

@article{duverger2024metrology,
  title={Metrology of microwave fields based on trap-loss spectroscopy with cold Rydberg atoms},
  author={Duverger, Romain and Bonnin, Alexis and Granier, Romain and Marolleau, Quentin and Blanchard, C{\'e}dric and Zahzam, Nassim and Bidel, Yannick and Cadoret, Malo and Bresson, Alexandre and Schwartz, Sylvain},
  journal={Physical Review Applied},
  volume={22},
  number={4},
  pages={044039},
  year={2024},
  publisher={APS}
}

@article{bohorquez2023reducing,
  title={Reducing Rydberg-state dc polarizability by microwave dressing},
  author={Bohorquez, JC and Chinnarasu, R and Isaacs, J and Booth, D and Beck, Matthew and McDermott, R and Saffman, M},
  journal={Physical Review A},
  volume={108},
  number={2},
  pages={022805},
  year={2023},
  publisher={APS}
}

@article{shi2017annulled,
  title={Annulled van der Waals interaction and fast Rydberg quantum gates},
  author={Shi, Xiao-Feng and Kennedy, TAB},
  journal={Physical Review A},
  volume={95},
  number={4},
  pages={043429},
  year={2017},
  publisher={APS}
}

@article{sevinccli2014microwave,
  title={Microwave control of Rydberg atom interactions},
  author={Sevin{\c{c}}li, Sevilay and Pohl, Thomas},
  journal={New Journal of Physics},
  volume={16},
  number={12},
  pages={123036},
  year={2014},
  publisher={IOP Publishing}
}

@article{young2021asymmetric,
  title={Asymmetric blockade and multiqubit gates via dipole-dipole interactions},
  author={Young, Jeremy T and Bienias, Przemyslaw and Belyansky, Ron and Kaufman, Adam M and Gorshkov, Alexey V},
  journal={Physical Review Letters},
  volume={127},
  number={12},
  pages={120501},
  year={2021},
  publisher={APS}
}

@article{kurdak2025enhancement,
  title={Enhancement of Rydberg Blockade via Microwave Dressing},
  author={Kurdak, Deniz and Banner, Patrick R and Li, Yaxin and Muleady, Sean R and Gorshkov, Alexey V and Rolston, SL and Porto, JV},
  journal={Physical Review Letters},
  volume={134},
  number={12},
  pages={123404},
  year={2025},
  publisher={APS}
}

@article{talashila2025determining,
  title={Determining angle of arrival of radio frequency fields using subwavelength, amplitude-only measurements of standing waves in a Rydberg atom sensor},
  author={Talashila, Rajavardhan and Watterson, William J and Moser, Benjamin L and Gordon, Joshua A and Artusio-Glimpse, Alexandra B and Prajapati, Nikunjkumar and Schlossberger, Noah and Simons, Matthew T and Holloway, Christopher L},
  journal={arXiv preprint arXiv:2502.09835},
  year={2025}
}

@article{elgee2025electrically,
  title={Electrically small Rydberg sensor for three-dimensional determination of radio-frequency k-vectors},
  author={Elgee, Peter K and Cox, Kevin C and Hill, Joshua C and Kunz, Paul D and Meyer, David H},
  journal={Physical Review Applied},
  volume={23},
  number={6},
  pages={064022},
  year={2025},
  publisher={APS}
}

@article{liu2023electric,
  title={Electric field measurement and application based on Rydberg atoms},
  author={Liu, Bang and Zhang, Lihua and Liu, Zongkai and Deng, Zian and Ding, Dongsheng and Shi, Baosen and Guo, Guangcan},
  journal={Electromagnetic Science},
  volume={1},
  number={2},
  pages={1--16},
  year={2023},
  publisher={CIE}
}

@article{yuan2023quantum,
  title={Quantum sensing of microwave electric fields based on Rydberg atoms},
  author={Yuan, Jinpeng and Yang, Wenguang and Jing, Mingyong and Zhang, Hao and Jiao, Yuechun and Li, Weibin and Zhang, Linjie and Xiao, Liantuan and Jia, Suotang},
  journal={Reports on Progress in Physics},
  volume={86},
  number={10},
  pages={106001},
  year={2023},
  publisher={IOP Publishing}
}

@article{kurdak2025high,
  title={High-Fidelity Microwave-Polarization Control in a Rydberg-Ensemble Experiment},
  author={Kurdak, Deniz and Li, Yaxin and Banner, Patrick R and Porto, JV and Rolston, SL},
  journal={arXiv preprint arXiv:2508.06820},
  year={2025}
}

@article{wang2023precise,
  title={Precise measurement of microwave polarization using a Rydberg atom-based mixer},
  author={Wang, Yuhan and Jia, Fengdong and Hao, Jianhai and Cui, Yue and Zhou, Fei and Liu, Xiubin and Mei, Jiong and Yu, Yonghong and Liu, Ya and Zhang, Jian and others},
  journal={Optics Express},
  volume={31},
  number={6},
  pages={10449--10457},
  year={2023},
  publisher={Optica Publishing Group}
}

@article{elgee2024complete,
  title={Complete three-dimensional vector polarimetry with a Rydberg-atom rf electrometer},
  author={Elgee, Peter K and Cox, Kevin C and Hill, Joshua C and Kunz, Paul D and Meyer, David H},
  journal={Physical Review Applied},
  volume={22},
  number={6},
  pages={064012},
  year={2024},
  publisher={APS}
}

@article{jiao2017atom,
  title={Atom-based radio-frequency field calibration and polarization measurement using cesium n DJ Floquet states},
  author={Jiao, Yuechun and Hao, Liping and Han, Xiaoxuan and Bai, Suying and Raithel, Georg and Zhao, Jianming and Jia, Suotang},
  journal={Physical Review Applied},
  volume={8},
  number={1},
  pages={014028},
  year={2017},
  publisher={APS}
}

@article{cloutman2025quantum,
  title={Quantum-enabled Rydberg atomic polarimetry of radio-frequency fields},
  author={Cloutman, Matthew and Chilcott, Matthew and Elliott, Alexander and Otto, J Susanne and Deb, Amita B and Kj{\ae}rgaard, Niels},
  journal={arXiv preprint arXiv:2503.17997},
  year={2025}
}

@article{sedlacek2013atom,
  title={Atom-based vector microwave electrometry using rubidium Rydberg atoms in a vapor cell},
  author={Sedlacek, JA and Schwettmann, A and K{\"u}bler, Harald and Shaffer, JP},
  journal={Physical review letters},
  volume={111},
  number={6},
  pages={063001},
  year={2013},
  publisher={APS}
}

@article{panja2024electric,
  title={Electric field control for experiments with atoms in Rydberg states},
  author={Panja, Aishik and Wang, Yupeng and Wang, Xinghan and Wang, Junjie and Subhankar, Sarthak and Liang, Qi-Yu},
  journal={AIP Advances},
  volume={14},
  number={12},
  year={2024},
  publisher={AIP Publishing}
}

@article{nill2024avalanche,
  title={Avalanche terahertz photon detection in a Rydberg tweezer array},
  author={Nill, Chris and Cabot, Albert and Trautmann, Arno and Gro{\ss}, Christian and Lesanovsky, Igor},
  journal={Physical Review Letters},
  volume={133},
  number={7},
  pages={073603},
  year={2024},
  publisher={APS}
}

@article{TuSQLelectrometer,
  title = {Approaching the standard quantum limit of a Rydberg-atom microwave electrometer},
  volume = {10},
  ISSN = {2375-2548},
  url = {http://dx.doi.org/10.1126/sciadv.ads0683},
  DOI = {10.1126/sciadv.ads0683},
  number = {51},
  journal = {Science Advances},
  publisher = {American Association for the Advancement of Science (AAAS)},
  author = {Tu,  Hai-Tao and Liao,  Kai-Yu and Wang,  Hong-Lei and Zhu,  Yi-Fei and Qiu,  Si-Yuan and Jiang,  Hao and Huang,  Wei and Bian,  Wu and Yan,  Hui and Zhu,  Shi-Liang},
  year = {2024},
  month = dec 
}

@misc{schlossberger2026MW,
      title={Resolving magnetic-sublevel structure in Rydberg Autler-Townes spectra with arbitrary RF polarization}, 
      author={Noah Schlossberger and Rajavardhan Talashila and Stone B. Oliver and Nikunjkumar Prajapati and William J. Watterson and Christopher L. Holloway},
      year={2026},
      eprint={2605.05466},
      archivePrefix={arXiv},
      primaryClass={physics.atom-ph},
      url={https://arxiv.org/abs/2605.05466}, 
}

\end{document}